\documentclass[10pt]{article}
\usepackage{soul}
\usepackage{authblk}
\usepackage{amsmath,amssymb,amsthm,amsfonts}
\usepackage{graphics,graphicx,epsfig}
\usepackage{wrapfig}
\usepackage{color}
\usepackage{multirow}
\usepackage{enumerate}
\usepackage{float}
\usepackage[hmargin=1in,vmargin=1in]{geometry}
\usepackage[backend=bibtex, sorting=none, bibstyle=ieee, style=numeric-comp,maxbibnames=99]{biblatex}
\usepackage{lipsum}
\usepackage[utf8]{inputenc}
\usepackage[symbol]{footmisc}
\usepackage{rotating}
\usepackage{appendix}
\usepackage[hmargin=1in,vmargin=1in]{geometry}
\usepackage{longtable}
\usepackage{caption}
\usepackage{url}
\usepackage{hyperref}
\numberwithin{equation}{section}

\newcounter{remark_cnt}
\stepcounter{remark_cnt}

\newenvironment{remark}{
\addtolength{\leftskip}{\parindent}
\noindent{\bf Remark \arabic{remark_cnt}:}\
}{
\par 
\vspace{10pt}
\stepcounter{remark_cnt}
}

\newcommand{\calD}{\mathcal{D}}%
\newcommand{\calF}{\mathcal{F}}%
\newcommand{\calS}{\mathcal{S}}%
\newcommand{\calV}{\mathcal{V}}%
\newcommand{\calW}{\mathcal{W}}%
\newcommand{\bfA}{{\bf A}}%
\newcommand{\bfb}{{\bf b}}\newcommand{\bfB}{{\bf B}}%
\newcommand{\bfC}{{\bf C}}%
\newcommand{\bfd}{{\bf d}}\newcommand{\bfD}{{\bf D}}%
\newcommand{\bfe}{{\bf e}}\newcommand{\bfE}{{\bf E}}%
\newcommand{\bfF}{{\bf F}}%
\newcommand{\bfG}{{\bf G}}%
\newcommand{\bfj}{{\bf j}}%
\newcommand{\bfL}{{\bf L}}%
\newcommand{\bfM}{{\bf M}}%
\newcommand{\bfn}{{\bf n}}%
\newcommand{\bfq}{{\bf q}}%
\newcommand{\bfr}{{\bf r}}\newcommand{\bfR}{{\bf R}}%
\newcommand{\bfS}{{\bf S}}%
\newcommand{\bft}{{\bf t}}\newcommand{\bfT}{{\bf T}}%
\newcommand{\bfu}{{\bf u}}\newcommand{\bfU}{{\bf U}}%
\newcommand{\bfv}{{\bf v}}\newcommand{\bfV}{{\bf V}}%
\newcommand{\bfW}{{\bf W}}%
\newcommand{\bfx}{{\bf x}}\newcommand{\bfX}{{\bf X}}%
\newcommand{\bfzeta}{\boldsymbol{\zeta}}%
\newcommand{\bfchi}{\boldsymbol{\chi}}%
\newfont{\tenbfit}{cmmib10}%
\newfont{\svnbfit}{cmmib8}%
\newfont{\tenbfsl}{cmbxti10}% <-- for idem tensor
\newfont{\mmit}{cmmi10}% scaled 1200
\newfont{\smit}{cmmi9}%  scaled 1200
\newfont{\bfMit}{cmmi5}%  scaled 1200
\newfont{\tenbbb}{msbm10}%
\newfont{\svnbbb}{msbm8}%
\newfont{\tenssit}{cmssqi8 at 10pt}%
\newfont{\svnssit}{cmssqi8 at 7pt}%
\newfont{\gothic}{eufm10}%
\newfont{\sgothic}{eufm7}%

\newcommand{\skw}{\hbox{\rm skw}\mskip3mu}
\newcommand{\sym}{\hbox{\rm sym}\mskip3mu}
\newcommand{\Tr}{\hbox{\rm tr}\mskip2mu}

\newcommand{\pards}[2]{\mbox{$\dfrac{\partial #1}{\partial {#2 }}$}}

\newcommand{\trans}{\mskip-2mu\scriptscriptstyle\top} %ELIOT I ADDED THE SPACING%

\newcommand{\Blj}{\mbox{$\Big[\kern-0.275em\Big[$}}
\newcommand{\Brj}{\mbox{$\Big]\kern-0.275em\Big]$}}
\newcommand{\zed}{{\bf 0}}

\newcommand{\Div}{\hbox{\rm Div}\mskip2mu}                                      %%%%%%
\newcommand{\B}{\text{B}}
\newcommand{\p}{\text{P}}                                                                                    %%%%%%

\newcommand{\X}{\bfX}

\newcommand{\y}{\bfchi}
\newcommand{\F}{\bfF}

\newcommand{\A}{\bfA}

\newcommand{\tendot}{\mskip-3mu:\mskip-2mu}

\newcommand{\Def}{\overset{\text{def}}{=}}

\newcommand{\ep}{\varepsilon}%
\newcommand{\thet}{\vartheta}%

\newcommand{\mat}{\text{\tiny R}}%
\newcommand{\T}{\bfT}
\newcommand{\Te}{\bfT^{\text{e}}}

\newcommand{\Me}{\bfM^\text{e}}

\newcommand{\Sym}{\text{sym}}

\newcommand{\Fe}{\bfF^\text{e}}

\newcommand{\Fei}{\bfF^{\text{e}-1}}
\newcommand{\Fet}{\bfF^{\text{e}\,\trans}}
\newcommand{\Feit}{\bfF^{\text{e}-\trans}}

\newcommand{\Le}{\bfL^\text{e}}

\newcommand{\Ee}{\bfE^{\text{e}}}           %%%%%%%%%%

\newcommand{\Ce}{\bfC^\text{e}}
\newcommand{\Cedot}{\dot{\bfC}^\text{e}}

\newcommand{\Ue}{\bfU^{\text{e}}}

\newcommand{\Fedot}{\dot{\bfF}^{\text{e}}}

\newcommand{\symz}{\text{sym}_0}
\newcommand{\Se}{\bfS^\text{e}}

\newcommand{\tmat}{\textbf{t}_{\text{R}}}

\newcommand{\xidot}{\mbox{$\dot{\xi}$}}
\newcommand{\dmatdot}{\dot{\textbf{d}}_{\text{R}}}
\newcommand{\xibardot}{\mbox{$\dot{\bar{\xi}}$}}

\newcommand{\Fc}{\bfF^c}
\newcommand{\Fg}{\bfF^\text{r}}
\newcommand{\Fi}{\bfF^{\text{c}}}
\newcommand{\Fidot}{\dot{\bfF}^{\text{c}}}
\newcommand{\Fm}{\bfF^{\text{m}}}

\newcommand{\Mme}{\bfM^{\text{e}}}

\newcommand{\Eei}{\text{E}_i^{\text{e}}}
\newcommand{\Cm}{\bfC^{\text{m}}}

\newcommand{\Fmdot}{\dot{\bfF}^{\text{m}}}
\newcommand{\Fgdot}{\dot{\bfF}^{\text{r}}}
\newcommand{\Fcdot}{\dot{\bfF}^{\text{c}}}
\newcommand{\Li}{\bfL^{\text{c}}}

\newcommand{\bfRe}{\bfR^\text{e}}
\newcommand{\bfRet}{\bfR^{\text{e}\trans}}
\newcommand{\mt}{\eta}
\newcommand{\Fmi}{\bfF^{\text{m}-1}}

\newcommand{\Fci}{\bfF^{\text{c}-1}}
\newcommand{\Fmt}{\bfF^{\text{m}\trans}}

\newcommand{\Cei}{\bfC^{\text{e}-1}}

\newcommand{\Fcit}{\bfF^{\text{c}-\trans}}

\newcommand{\Fii}{\bfF^{\text{c}-1}}

\newcommand{\Ji}{J^{\text{c}}}

\newcommand{\Je}{J^{\text{e}}}
\newcommand{\Jei}{J^{\text{e}-1}}
\newcommand{\Jm}{J^{\text{m}}}  
\newcommand{\Jg}{J^{\text{r}}}  
\newcommand{\Jgi}{J^{\text{r}-1}}  
\newcommand{\Jc}{J^{\text{c}}}  
\newcommand{\Jci}{J^{\text{c}-1}}

\newcommand{\Lm}{\bfL^{\text{m}}}   
\newcommand{\Lg}{\bfL^{\text{r}}} 
\newcommand{\Lc}{\bfL^{\text{c}}} 
\newcommand{\Ld}{\bfL^{\text{d}}} 
\newcommand{\Di}{\bfD^{\text{c}}}
\newcommand{\Dm}{\bfD^{\text{m}}}   
\newcommand{\Dg}{\bfD^{\text{r}}}  
\newcommand{\Dc}{\bfD^{\text{c}}}

\newcommand{\Wc}{\bfW^{\text{c}}}  
\newcommand{\psimech}{\psi_\mat^{\text{m}}}

\newcommand{\cdotr}{\dot{c}}
\newcommand{\ccr}{c}
\newcommand{\ccro}{c_{0}}
\newcommand{\ccrmax}{c_{\text{max}}}
\newcommand{\cbar}{\bar{c}}
\newcommand{\xidotr}{\dot{\xi}}
\newcommand{\xir}{\xi}

\newcommand{\qr}{q_\mat}
\newcommand{\imat}{\textbf{i}_\mat}
\newcommand{\emat}{\textbf{e}_\mat}
\newcommand{\dmat}{\textbf{d}_\mat}
\newcommand{\mue}{\mu^{\text{e}}}

\newcommand{\Ng}{\textbf{N}^{\text{r}}}

\newcommand{\psic}{\psi_\mat^c}

\newcommand{\muM}{\mu}
\newcommand{\muxi}{\mu^{\xi}}

\newcommand{\xibar}{\bar{\xi}}

\newcommand{\cc}{c}

\newcommand{\cco}{c_0}

\newcommand{\lambdaei}{\lambda_i^{\text{e}}}

\newcommand{\ximax}{\xi^\text{max}}

\newcommand{\muxiz}{\mu_0^\xi}

\newcommand{\dam}{\text{d}}
\newcommand{\damdot}{\dot{\text{d}}}
\newcommand{\stressdis}{\varpi_{\text{dis}}}
\newcommand{\stressen}{\varpi_{\text{en}}}

\newcommand{\epsgdot}{\dot{\epsilon}^{\text{r}}}
\newcommand{\Sg}{\textbf{S}^{\text{r}}}
\newcommand{\mg}{\textbf{m}_{\text{r}}}

\newcommand{\muo}{\mu_{0}}

\newcommand{\psimechpos}{\psi_\mat^{\text{m+}}}
\newcommand{\psimechneg}{\psi_\mat^{\text{m-}}}

\newcommand{\Eeone}{\text{E}_1^{\text{e}}}
\newcommand{\Eetwo}{\text{E}_2^{\text{e}}}
\newcommand{\Eethree}{\text{E}_3^{\text{e}}}
\newcommand{\psicr}{\psi^{*}}
\newcommand{\psicrgb}{\psi_\text{gb}^{*}}
\newcommand{\psicrb}{\psi_\text{bulk}^{*}}
\newcommand{\Mmepos}{\textbf{M}^{\text{e+}}}
\newcommand{\Mmeneg}{\textbf{M}^{\text{e-}}}

\newcommand\Trule{\rule{0pt}{2.6ex}}
\newcommand\Brule{\rule[-1.2ex]{0pt}{0pt}}

\bibliography{references}

\begin{document}

\title{
\textbf{
A continuum electro-chemo-mechanical gradient theory coupled with damage: Application to Li-metal filament growth in all-solid-state batteries.
}
}

\author{Donald Bistri and Claudio V. Di Leo\thanks{Corresponding Author: E-Mail Address: cvdileo@gatech.edu}
}
\affil{School of Aerospace Engineering, Georgia Institute of Technology}
\date{}
\maketitle

%%%%%%%%%%%%%%%%%%%%%%%%%%%%%%%%%%%%%%%%%%%%%%%%%
%%%%%%%%%%%%%%%%%%%%%%%%%%%%%%%%%%%%%%%%%%%%%%%%%
%%%%%%%%%%%%%%%%%%%%%%%%%%%%%%%%%%%%%%%%%%%%%%%%%

\section*{Abstract}

We formulate a thermodynamically-consistent electro-chemo-mechanical gradient theory which couples electrochemical reactions with mechanical deformation and damage in solids. 
The framework models both species transport across the solid host due to diffusion/migration mechanisms and concurrent electrochemical reaction at damaged zones within the solid host, where ionic species are reduced to form a new compound.
The theory is fully-coupled in nature with electrodeposition impacting mechanical deformation, stress generation and subsequent damage of the solid host. 
Conversely, electrodeposition kinetics are affected by mechanical stresses through a thermodynamically-consistent, physically motivated driving force that distinguishes the role of chemical, electrical and mechanical contributions.
The framework additionally captures the interplay between growth-induced fracture of the solid host and electrodeposition of a new material inside cracks by tracking the damage and extent of electrodeposition using separate phase-field variables.

While the framework is general in nature, we specialize it towards a critical problem of relevance to commercialization of next-generation all-solid-state batteries, namely the phenomenon of Li-metal filament growth across a solid-state electrolyte. 
We specialize on a Li-metal - Li$_{7}$La$_{3}$Zr$_{2}$O$_{12}$ (LLZO) system and demonstrate the ability of the framework to capture both intergranular and transgranular crack and Li-filament growth mechanisms, both of which have been experimentally observed. 
In addition, we elucidate the manner in which mechanical confinement in solid-state batteries plays an important role in the resulting crack/electrodeposition morphology.
In modeling this Li/LLZO system, we demonstrate the manner in which our theoretical framework can elucidate the critical coupling between mechanics and electrodeposition kinetics and its role in dictating Li-filament growth.
Beyond this application, the theoretical framework should serve useful in a number of engineering problems of relevance in which electrochemical reactions take place within a
damage zone, leading to deposition of new material at these locations.

\vspace{0.1in}

\noindent
{\bf Keywords:} Continuum electro-chemo-mechanics; 
Dendrite growth; 
Phase-field formulation; 
Energy storage;
Solid-State Battery.

\newpage
%%%%%%%%%%%%%%%%%%%%%%%%%%%%%%%%%%%%%%%%%%%%%%%%
%%%%%%%%%%%%%%%%%%%%%%%%%%%%%%%%%%%%%%%%%%%%%%%%
%%%%%%%%%%%%%%%%%%%%%%%%%%%%%%%%%%%%%%%%%%%%%%%%
%
\section{Introduction}
\label{sect:intro}

Solid-state batteries (SSBs) present a promising technology for next-generation energy storage systems. In recent years, commercialization of SSBs has attracted significant research attention owing to their promise of superior performance compared to conventional Lithium-ion batteries (LIBs). 
Among their many advantages, SSB architectures can enable for increased current density ($\approx$ 3860 mAh/g), improved safety, and a wider electro-chemical window (0-5V) - in turn enabling for coupling of Li metal anodes with high voltage cathode materials \cite{yao2016high,li2015solid,zhu2015origin,yang2017protected,roth2012electrolytes}. 
From a safety standpoint, use of inorganic solid-state electrolytes (SSEs) immediately alleviates potential hazards associated with ignition of liquid electrolytes commonly used in LIBs.
In addition, the solid-state nature of the electrolyte material has the potential to alleviate failure mechanisms associated with dendrite growth. 
However, successful design and operation of SSBs is hampered by several chemo-mechanical challenges across its constituents \cite{wang2019electro,bistri2021modeling,zhang2020review,bistri2021modeling_ecs,tippens2019visualizing,bistri2021roledamage}, the most critical one associated with metal filament growth across the SSE microstructure, eventually leading to short-circuit.

Unlike the process of dendrite growth in liquid electrolyte systems, growth of metal filaments across the SSE in a solid-state architecture depends on fundamentally different mechanics. 
Owing to the solid-state nature of the electrolyte, the SSE microstructure is hypothesized to play a fundamental role. 
Li-metal filaments can initiate at imperfections or heterogeneities of the metal/SSE interface such as voids, cracks, and grain boundaries. Once initiated, these filaments propagate through the SSE microstructure by fracturing the latter and creating fresh sites for continuing Li-metal deposition and filament growth. 
In this view, {\it mechanical fracture and Li-metal filament growth are intrinsically coupled}.
Eventually, metal filaments may pierce through the entire SSE, reaching the cathode and leading to short-circuit of the battery \cite{ren2015direct,cheng2017intergranular,porz2017mechanism,ning2021visualizing}.
It is thus critical to understand from both an experimental and modeling perspective the interplay between electrodeposition at the Li-metal interface and the SSE microstructure. In particular, the manner in which defects, rate of Li deposition, local stress state, and mechanical behavior (including fracture) of both Li-metal and SSE couple to govern the onset and evolution of Li-metal filaments in SSB architectures \cite{lewis2021linking,shen2018effect,yan2022does,anand2019elastic,lepage2019lithium,singh2022li6ps5cl}.

Filament growth has been experimentally observed to occur above a critical current density in both compliant \cite{ning2021visualizing,hao20213d,lewis2022role,nagao2013situ} and stiff oxide-based\cite{cheng2017intergranular,ren2015direct,porz2017mechanism,hao2021tracking,kazyak2020li,krauskopf2019toward} electrolytes.
These experimental observations counter previous modeling predictions \cite{monroe2005impact} that claimed filament growth could be suppressed - and replaced with stable Li-metal deposition - provided the shear modulus of the SSE is at least twice as large as that of Li-metal. 
Advanced characterization techniques including tomographic imaging and spectroscopy have  revealed propagation of Li-metal filaments across the SSE microstructure to occur either i) transgranularly, in the form of a dominant Li-filament, typical for single-crystal electrolytes \cite{ning2021visualizing,hao2021tracking,swamy2018lithium,porz2017mechanism} or ii) intergranularly, through the grain boundaries, typical for polycrystalline SSEs \cite{cheng2017intergranular,ren2015direct,hao20213d,heo2021short,xu2022structural}. 
Recent studies have additionally hypothesized the potential for isolated Li-metal deposits to form within the bulk SSE microstructure due to the favorable SSE electronic properties and  presence of trapped electrons. Upon reaction with Li-ions, these trapped electrons may produce isolated Li-metal deposits, seeding subsequent  growth of metal filaments across the SSE microstructure
\cite{han2019high,tian2019interfacial}.

Aside from experimental observations, several modeling frameworks have been proposed in the literature to address different aspects of the nucleation and propagation of Li-metal filaments in SSEs. 
Related to nucleation of Li-metal filaments, linear 
perturbation analysis has 
been extensively employed in 
several models \cite{barai2016effect,mcmeeking2019metal,ahmad2017stability,barai2018impact,natsiavas2016effect,ahmad2017role}.
Starting with a sinusoidal interface perturbation, these models have investigated the stability of electrodeposition at the Li-metal/SSE interface with variation in perturbation amplitude, applied current  density and state of pre-stress.  
Consistent with experimental observations, growth of Li-metal filaments (e.g. unstable electrodeposition) was shown to occur irrespective of the SSE stiffness provided the  applied current density and perturbation wavelength are sufficiently high.
To study the potential for propagation of existing Li-metal filaments, a second class of models, treating metal protrusions as pressurized cracks under a linear elastic fracture mechanics framework was later developed \cite{porz2017mechanism,klinsmann2019dendritic,bucci2019modeling,ansell1986chemical,armstrong1974breakdown,feldman1982initiation}.
These works investigated the role that defect size, surface
resistance and tip overpotential play on the propensity for Li-metal filaments to fracture the SSE and grow across the solid conductor.
While clearly important, the two classes of models discussed above are limited to either modeling the nucleation of  Li-filaments or the conditions under which an existing filament might propagate. 
They can not capture the dynamic evolution of Li-metal filaments, which is the focus of this work.

Recently, mature continuum chemo-mechanical frameworks have been developed to model the complex electro-chemo-mechanical processes that govern non-uniform electrodeposition at a Li-metal/SSE interface in the presence of imperfections. 
Narayan and Anand \cite{narayan2020modeling}  propose a thermodynamically-consistent continuum theory to model non-uniform platting/stripping kinetics at the Li-anode coupled with elastic-viscoplastic deformation of Li-metal.
Given the complexity of modeling freshly-deposited Li-layers from a finite element standpoint, the authors propose an analogous mechanical problem of swelling and de-swelling of a thin ``interphase layer" at the Li-metal/SSE interface to reproduce the electro-chemical process of plating/stripping. 
The framework provides important insights on the role that both geometric imperfections and chemical impurities at the metal/SSE interface play on the evolution of Li-metal filaments and concomitant deformation and fracture of the associated metal electrode-solid electrolyte materials.
However, the framework is limited in that it does not directly model the ongoing electrochemical reaction kinetics driving the growth of Li-metal filaments and can not subsequently predict the morphology and evolution of the resulting Li-filaments. 
The work of Shishvan et al. \cite{shishvan2020dendrites,shishvan2020growth} presents one of the first attempts to model the evolution of Li-metal filaments across the SSE beyond the idealization of filaments as pressurized cracks.
In \cite{shishvan2020dendrites}, the authors propose an alternative mechanism for initiation of metal filaments, treating metal protrusions as thick-edge dislocations and demonstrate the capacity of the framework in \cite{shishvan2020growth} to predict the experimentally reported critical-current density and growth rate of Li-deposits across SSEs.

Phase-field models have also been extensively developed to capture  Li-metal filament growth in both liquid and solid-state battery architectures.
Purely electro-chemical non-linear phase-field models have been developed and shown to successfully predict the evolution and morphology of Li dendrites in liquid-electrolyte systems under varying applied electric potentials \cite{chen2015modulation,shibuta2007phase,ely2014phase,hong2018phase,liang2012nonlinear}.
Recently, phase-field based models have been extended to solid-state architectures - including both inorganic and polymer-based electrolytes - to incorporate the role of mechanics and deformation of Li-metal due to the confining SSE on electrodeposition kinetics \cite{tantratian2021unraveling,tian2019interfacial,shen2021does,ren2020inhibit}. 
These works present the first attempts at capturing the evolution and morphology of Li-metal filaments across a solid conductor, while simultaneously accounting for the presence of microstructural inhomogeneities in the solid-electrolyte \cite{tantratian2021unraveling}. 
However, these models do not capture the fundamental coupling between fracture of the SSE and electrodeposition. 
Li-metal filaments growing in solid electrolytes, unlike dendrites in their liquid counterparts, are constrained by a solid continuum and must overcome the associated mechanical resistance by fracturing the solid host.
Only upon fracture of the solid conductor, and subsequent creation of an empty deposition site, may Li-metal filaments proceed to grow by plating the newly fractured space.
Recent work by Fincher et al. \cite{fincher2022controlling} experimentally demonstrates this phenomenon. There, the authors apply an external, experimentally controlled stress field on a solid-state electrolyte through which a Li-filament is growing. 
By controlling the stress field, Fincher and co-workers change the preferred fracture direction of the underlying SSE and demonstrate that this in turn changes the direction of Li-filament growth correspondingly. 
This work provides direct experimental evidence for the coupling between SSE fracture and Li-metal filament growth.

As summarized above, there is a need for a fully-coupled, electro-chemo-mechanical framework which captures: i) the nucleation and evolution of Li-metal filaments, ii) the concurrent fracture of the solid host, and iii) the coupling between mechanical deformation, stress and Li-metal electrodeposition kinetics. 
Of particular importance in such a framework is also the ability to model the role that heterogeneities in the SSE microstructure (i.e. pores, cracks, grain boundaries) play on the coupled fracture behavior of the SSE and concurrent Li-metal filament growth due to electrodeposition. 
The purpose of this work is to present such a framework.

We develop here a thermodynamically consistent, phase-field electro-chemo-mechanical framework for modeling concurrent diffusion, reaction, deformation and damage in solids.
While the framework is general in nature and can be applied across a number of engineering problems (i.e. electro-chemically active polymers \cite{mao2018theory,narayan2021coupled,narayan2022coupled,cha2014mechanics}, oxidation and corrosion \cite{nguyen2017modeling,konica2020thermodynamically,cui2021phase,loeffel2011chemo,ammar2009finite}, electro-refining \cite{lin2022mechano}), we specialize it towards modeling the onset and evolution of Li-metal filaments in all-solid-state batteries.
In particular, our framework includes the following unique features distinguishing it from previous work in the literature:
\begin{itemize}
    \item 
    The theoretical framework captures concurrent charged species diffusion/migration and reaction, coupled to mechanical deformation and damage of the solid host.
    Currently, continuum electro-chemo-mechanics theories for energy storage materials in the literature have been largely developed for diffusion of a conserved species across the host (i.e. no chemical reactions) \cite{anand2012cahn,di2014cahn,di2015diffusion,bower2011finite,bower2012simple}, with some recent efforts capturing diffusion-reaction in solids \cite{loeffel2013modeling,zhao2019diffusion,afshar2021thermodynamically}.
    While rigorous, current diffusion-reaction-deformation frameworks do not consider transport of charged species across the host in the presence of electric field, and thus are limited to modeling chemical rather than electro-chemical reactions.
    Continuum mechanics formulations for transport of charged species have been developed \cite{narayan2021coupled,narayan2022coupled,zhao2022phase}, but in turn do not consider electrochemical reactions and apply only to a conserved diffusing species.
    This work provides a thermodynamically consistent electro-chemo-mechanical framework, which concurrently accounts for diffusion and reactions within the host material with the potential for reaction-induced fracture. 
    %
    %%%
    \item 
    The theory captures the coupling between fracture of the solid host and electrodeposition of a new material inside cracks by modeling the evolution of cracks and material deposition using distinct phase-field variables.
    Importantly, the theory presents the first framework coupling a phase-field reaction model for Li-metal filament growth with a phase-field damage model for electrodeposition-induced fracture of the solid conductor.
    Critically, this enables one to model the manner in which mechanical fracture of the SSE is intrinsically coupled to electrodeposition and growth of Li-metal filaments. We demonstrate how this coupling can arise both through a thermodynamically consistent reaction driving force, and/or through a damage dependent formulation of the reaction kinetics. 
    %
    %%%
    \item 
    The theoretical framework, and accompanying numerical implementation, allows one to model the role of microstructure heterogeneities (i.e. voids, cracks, grain boundaries) with varying chemo-mechanical properties on concurrent fracture of the SSE and electrochemical growth of Li-filaments. 
    As such, the model may predict the manner in which various microstructural features enhance or suppress Li-filament growth and subsequently guide the development of microstructures for minimizing filament growth induced degradation. 
\end{itemize}

To demonstrate the relevance and use of the proposed theoretical framework, we specialize it towards modeling the growth of Li-metal filaments in an inorganic Lithium Lanthanum Zirconium Oxide (LLZO) electrolyte, specifically  Li$_{7}$La$_{3}$Zr$_{2}$O$_{12}$. 
LLZO is a promising SSE candidate due to its high ionic conductivity, high mechanical stiffness and good stability against Li-metal. 
We explore the experimentally observed manner in which Li-metal filaments may grow across the LLZO electrolyte either i) transgranularly, as a single dominant Li-metal filament propagating across the conductor \cite{porz2017mechanism}, and/or ii) intergranularly, across the mechanically weaker grain boundaries \cite{hao20213d,cheng2017intergranular,ren2015direct}.
In particular, owing to the coupling between mechanical stresses, fracture and electrodeposition kinetics developed in this framework, we investigate the manner in which mechanically weaker grain boundaries with lower fracture energy dictate the transgranular versus intergranular nature of the resulting Li-metal filaments. 
Finally, using our framework, we provide insights on the manner in which mechanical confinement and stresses dictate the advancement of cracks versus Li-metal deposits across the solid electrolyte. 
In particular, whether the crack front propagates significantly ahead of the Li-metal filaments as has been experimentally observed \cite{ning2021visualizing,hao20213d}.

The paper is organized as follows.
In Sect.  \ref{sect:Bal_Eqns} we introduce mass balance, charge balance and electrostatics equations and define the physically motivated phase-field parameters, $\xibar$, governing the extent of electrodeposition, and $\dam$, governing the extent of damage in the solid host.
Kinematics are developed in Sect.~\ref{sect:Kinematics}. 
The remaining governing laws are developed in Sect.~\ref{sect:bal_laws} using the principle of virtual power and the first and second laws of thermodynamics.
The constitutive theory is presented in Sect. \ref{Const_Theory} and summarized in its general form in Sect. \ref{sect:summ_general}. In Sect.~\ref{theory_spec}, we present a specialization of the theory towards modeling growth of Li-metal filaments in a LLZO solid-state electrolyte and summarize the governing partial differential equations and initial/boundary
conditions in Sect. \ref{sect:PDE-sec}.
Numerical simulations are presented in Sect. \ref{num_sims}. 
First, in Sect.~\ref{transg_growth}, we model growth of a single Li-metal filament across a LLZO electrolyte. Through this model, we demonstrate the salient features of our theoretical framework and explore the role that mechanical confinement and stress play on the evolution of cracks and Li-metal filament growth across the solid electrolyte.
In Sect.~\ref{dam_mech_kinetics}, we illustrate the manner in which different microstructural features, in particular grain boundaries, may be incorporated into the modeling framework. 
Within this model of a LLZO electrolyte with resolved grain boundaries, we elucidate the manner in which mechanical stresses, damage and electrodeposition are coupled and how different coupling mechanisms result in different fracture and Li-filament morphologies.
Finally, we demonstrate the manner in which this framework directly couples mechanical properties, such as grain boundary fracture energy, with the resulting Li-filaments growth pattern. 
We close with concluding remarks in Sect. \ref{conclusion}.

%%%%%%%%%%%%%%%%%%%%%%%%%%%%%%%%%%%%%%%%%%%%%%%%%%%%%%%%%%%%%%%%%%%%%%%%%%%
%%%%%%%%%%%%%%%%%%%%%%%%%%%%%%%%%%%%%%%%%%%%%%%%%%%%%%%%%%%%%%%%%%%%%%%%%%%
%%%%%%%%%%%%%%%%%%%%%%%%%%%%%%%%%%%%%%%%%%%%%%%%%%%%%%%%%%%%%%%%%%%%%%%%%%%

\section{Phase-field formulation and balance equations}
\label{sect:Bal_Eqns}

\begin{figure}[h]
    \centering
    \includegraphics[width=6in]{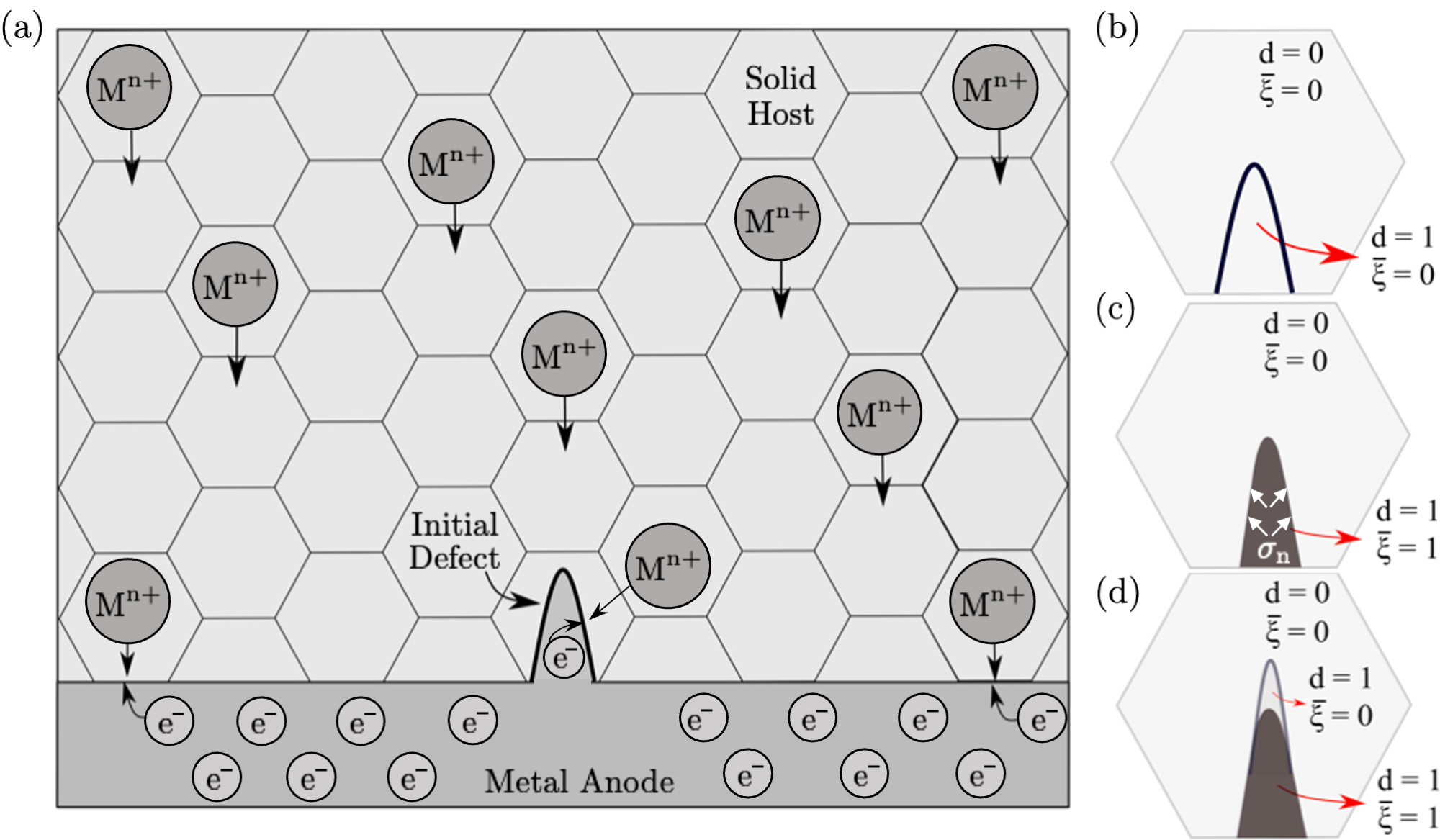}
    \caption{ (a) Schematic of a metal electrode-solid conductor assembly. 
    (b) Illustrates a pre-existing defect ($\xibar=0,\dam=1$) at the solid host-metal electrode interface prior to metal deposition inside the crack. 
    (c) Illustrates deposition of metal inside the imperfection ($\xibar = 1, \dam=1$), leading in turn to a build up in normal stresses, $\sigma_\text{n}$, at the crack surfaces. 
    (d) Build-up in normal stresses on the crack surface eventually causes the solid host to damage further ahead of the crack tip, creating a new damage zone ($\xibar=0,\dam=1$), which can accommodate subsequent deposition of metal inside the solid host.
    }
    \label{fig:sse_schematic}
\end{figure}

This section details the conservation laws for electrodeposition phenomena within a solid ion-conducting host. The problem
is shown schematically in Fig.~\ref{fig:sse_schematic}(a), where we illustrate a solid conductor (SSE) adjacent to an electrode composed of a metallic compound “M”. The solid conductor/metal electrode interface may contain a number
of defects (i.e. pores, voids, cracks), which serve as vacant sites for the metallic compound to fill and subsequently deposit inside the solid host (c.f. \cite{porz2017mechanism,klinsmann2019dendritic,bucci2019modeling}).
Consider now a simple and general electrodeposition reaction. Cations, M$^{n+}$ conduct across the solid host towards the conductor/metal interface, where they react with electrons, e$^{-}$ to deposit M-atoms as follows
\begin{equation}
\text{M}^{n+} + \text{ne}^{-}\rightarrow \text{M}.
\label{Li_reac}
\end{equation}
The electrodeposition of M-atoms (i.e. metallic compound) is treated in a phase-field sense and taken to occur over a diffuse boundary, with the extent of reaction tracked via a normalized phase-field parameter
\begin{equation}
    \xibar = \xir/\ximax \in [0,1].
\end{equation}
Here, $\xir$ denotes the moles of electrodeposited species per unit reference volume, with $\ximax$ the maximum amount of the metallic compound that can be electrodeposited at the reaction site.
Physically, a state of $\xibar = 0$ denotes the absence of metallic compound or “void”, while a state of $\xibar$ = 1 denotes the fully deposited metallic compound.
Naturally then, intermediate values of $0 < \xibar <1$ represent the reaction zone.

In this physical interpretation of the phase field variable, $\xibar$, we  restrict the formulation such that \textit{electrodeposition inside the solid host may occur only within voids }(i.e. cracks, pores), which either form through mechanical loading or exist inherently in the microstructure. This interpretation conforms with experimental observations that for metal deposits to grow through a solid host, they must first overcome mechanical resistance by fracturing the solid to create the necessary vacant space to accommodate plating (c.f. Fincher et al. \cite{fincher2022controlling} and Ren et al. \cite{ren2019microstructure}).

We now introduce a phase-field damage variable, $\dam(\bfX,t)$ to describe any state of imperfections across the solid host, either pre-existing (i.e. pores, voids) or mechanically induced (crack formation) 
\begin{equation*}
    \dam(\bfX,t) \in [0,1].
\end{equation*}
\label{dam_eq1}
Here, a state of $\dam = 0 $ represents the intact material, while a state of $\dam = 1$ represents the fully-fractured solid host. Intermediate values, $ 0 < \dam < 1$ naturally denote then a partially fractured material. Additionally, we assume microstructural changes leading to fracture to be irreversible and require damage to grow monotonically such that 
\begin{equation}
    \damdot(\bfX,t)> 0.
\label{dam_dot}
\end{equation}

With a focus on the crack/electrodeposition interplay, we restrict ourselves such that {\it electrodeposition may only occur within damaged regions of the SSE}. That is,
\begin{equation}
    \xibar > 0 \quad \text{only if} \quad \dam > 0.
\end{equation}
This is shown schematically in Fig. \ref{fig:sse_schematic}(b-d). 
Fig.~\ref{fig:sse_schematic}(b) illustrates a microstructural defect in the form of a void or crack (region of $\dam=1$) at the metal/SSE interface, which provides the necessary vacant space for metal to deposit.
With continuous deposition, the metallic compound progressively covers the entire area of the crack ($\dam = 1$, $\xibar = 1$), as illustrated in Fig. \ref{fig:sse_schematic}(c).
This in turn induces a build-up in stresses acting on the crack's surface, large enough to eventually overcome the fracture toughness of the solid conductor and incur further damage on the solid host,  Fig. \ref{fig:sse_schematic}(d). 
The newly formed damaged region ($\dam = 1, \xibar = 0$) then provides the additional vacant site for the metal to deposit further into the solid electrolyte.
This synergistic mechanism sustains the growth of Li-metal filament across the solid host. 
Eventually, the metallic protrusions fracture and transverse the entire span of the solid conductor, at which point the battery short-circuits.  

%%%
%%%
%%%
\subsection{Mass Balance}
\label{sect:Mass_Bal}

Considering the electrodeposition reaction \eqref{Li_reac}, conservation of mass for the mobile ionic species transporting across the solid host may then be written as a diffusion-reaction equation of the form
\begin{equation}
    \cdotr = -\Div \bfj_\mat - \xidotr. 
    \label{mass_bal1}
\end{equation}
Here, $\cc(\bfX,t)$ denotes the number of moles of the mobile ionic species per unit reference volume, while $\xir(\bfX,t)$ denotes the number of moles of the electrodeposited species per unit reference volume with $\xidotr$ the electrodeposition rate. 
Consistent with the discussion in Sect. \ref{sect:Bal_Eqns}, when the extent of reaction $\xibar = 1$, the reaction has led to consumption of $\ximax$ moles of the diffusing species.
Additionally, $\bfj_\mat (\X,t)$ represents the referential flux of the mobile ionic species transporting across the solid host, which must be constitutively prescribed. 
Note here that the reaction-diffusion equation (\ref{mass_bal1}) encapsulates both transport of ionic species across the solid host via the $\Div \bfj_\mat$ term and its subsequent consumption leading to formation of the newly electrodeposited metallic compound via the $\xidotr$ term.

%%%
%%%
%%%
\subsection{Charge Balance}
\label{sect:charge_balance}

The net charge per unit reference volume of the solid conducting host, $\qr$ is given by (c.f. Narayan and Anand \cite{narayan2021coupled} and Li and Monroe. \cite{li2019dendrite})
\begin{equation}
    \qr = F(\cc - \cco),
    \label{char_balance}
\end{equation}
where $\cco$ represents the initial (i.e. bulk) concentration per unit reference volume of the mobile ionic species transporting across the solid host, and $F$ is the Faraday constant.
Note here that in writing Eqn. \eqref{char_balance}, we restrict our attention to monovalent cationic species and additionally assume negatively charged species in the solid host to be immobile. This restricts the
mass flux across the solid host to be carried only by the mobile cations. As a result, any local excess charge density across the solid host can be induced only by a change in concentration of the mobile cationic species from the equilibrium bulk concentration as they transport across the solid conductor.

Defining a referential current density, 
\begin{equation}
    \imat \Def F\bfj_\mat
\end{equation}
taking the time derivative of \eqref{char_balance}, and making use of mass balance \eqref{mass_bal1}, one may write for later use
\begin{equation}
    \dot{\qr} = -\Div \imat - F\xidotr.
     \label{charge_balance_v2}
\end{equation}

%%%
%%%
%%%
\subsection{Electrostatics}
\label{sect:Electrostatics}

Under electrostatic conditions, for a polarizable material we require that the two governing Maxwell equations are satisfied. 
The first equation, Faraday's law, requires the referential electric field, $\emat(\bfX,t)$ to obey the following relationship
\begin{equation}
    \text{curl} \, \emat = 0.
    \label{Faraday_law}
\end{equation}
One can automatically satisfy Eqn. \eqref{Faraday_law} by representing the electric field, $\emat(\bfX,t)$ as the gradient of an electrostatic potential, $\phi(\bfX,t)$ such that
\begin{equation}
    \emat(\bfX,t) = - \nabla \phi(\bfX,t).
    \label{Faraday_law2}
\end{equation}
The second equation of electrostatics is given by Gauss’s law, and relates the electric displacement, $\dmat(\bfX,t)$ to the net charge per unit reference volume, $\qr$ as follows
\begin{equation}
    \text{Div} \, \dmat = \qr.
    \label{gauss_law}
\end{equation}
In conjunction, eqs. \eqref{Faraday_law} and \eqref{gauss_law} constitute the governing equations for electrostatics in the solid host.

%%%%%%%%%%%%%%%%%%%%%%%%%%%%%%%%%%%%%%%%%%%%%%%%%%%%%%%%%%%%%%%%%%%%%%%%%%%
%%%%%%%%%%%%%%%%%%%%%%%%%%%%%%%%%%%%%%%%%%%%%%%%%%%%%%%%%%%%%%%%%%%%%%%%%%%
%%%%%%%%%%%%%%%%%%%%%%%%%%%%%%%%%%%%%%%%%%%%%%%%%%%%%%%%%%%%%%%%%%%%%%%%%%%

\section{Kinematics}
\label{sect:Kinematics}

We introduce here the kinematic formulation for description of a deformation resulting from coupled species transport, distortion due to electrodeposition of a new material inside the solid host, and mechanics.
Consider a macroscopically homogeneous body $\mathcal{B}$ with the region of space it occupies in a fixed reference configuration and let $\bfX$ denote an arbitrary material point of $\mathcal{B}$. The motion of $\mathcal{B}$ is then a smooth one-to-one mapping $ \bfx = \y(\bfX,t) $ with deformation gradient, velocity, and velocity gradient given by\footnote{Notation: We use standard
notation of modern continuum mechanics (c.f. Gurtin et al. \cite{gurtin2010mechanics}). Particularly: $\nabla$ and
Div denote the gradient and divergence operators with respect to the material point $\bfX$ in the reference configuration; while grad and div operate on the point $\bfx=\bfchi(\bfX,t)$ in the  deformed body; a superposed dot denotes the material time-derivative. 
Throughout, we write $\bfF^{e-1} = (\bfF^e){}^{-1}$, $\bfF^{e-\trans}=(\bfF^e)^{-\trans}$, etc. We also write $\Tr\A$, $\sym\A$, $\skw\A$, $\A_0$, and $\symz\A$ respectively, for the trace, symmetric, skew, deviatoric, and symmetric-deviatoric parts of a tensor $\A$. 
Finally, the inner product of tensors $\A$ and $\bfB$ is denoted by $\A \tendot \bfB$, and the magnitude of $\A$ by $|\A|=\sqrt{\A \tendot \A}$}.
\begin{equation}
\bfF = \nabla \y, \qquad \bfv = \dot{\y}, \qquad \bfL = \text{grad}\,\bfv = \dot{\bfF}\bfF^{-1}.
\label{kin1}
\end{equation}
The theory builds on a multiplicative decomposition of the deformation gradient,
\begin{equation}
\bfF = \bfF^\text{mechanical} \bfF^\text{chemical} = \Fm\Fc.
\label{mult_dec}
\end{equation}
Here
\begin{itemize}
%
%%%
\item[(i)]  $\Fc(\bfX)$ represents the concurrent local distortion of the material neighborhood of $\X$ due to chemical phenomena including: i) species transport across the solid host electrolyte, and ii) distortion due to electrodeposition of a new material inside the solid host at the reaction sites. 
%
%%%
\item [(ii)]
$\Fm$ represents the distortion due to macroscopic stresses and may include local irreversible plastic deformation caused by inelastic mechanisms such as dislocation motion, and the subsequent elastic stretching and rotation of the inelastically deformed material neighborhood. 
\end{itemize}

% \vspace{10pt}
\begin{remark}
In modeling local distortions due to electrodeposition through the tensor $\Fc$, the chemical deformation tensor acts as a surrogate for modeling the continuous creation of new material and ensuing distortions, without explicitly accounting for the temporal evolution of the domain. Similar continuum treatments with a focus on growth of electrochemical interphases and growing matter in living systems can be found in the works of Narayan and Anand \cite{narayan2020modeling}, Tantratian et al. \cite{tantratian2021unraveling} and Kuhl \cite{kuhl2014growing}.
\qed
\end{remark}

We refer to $\Fm$ and $\Fc$ as the mechanical and chemical distortions respectively. 
The volume ratio is given by
\begin{equation}
J \Def \text{det}\,\bfF >0,
\label{vol_constraint}
\end{equation}
and using \eqref{mult_dec}, 
\begin{equation}
J=\Jm \Jc,
\quad \text{with} \quad
\Jm \Def \text{det}\,\Fm>0,
\quad \text{and} \quad 
\Jc \Def\text{det}\,\Fc>0,
\label{JeJg}
\end{equation}
such that $\Fm$ and $\Fc$ are invertible. The right and left polar decomposition  of $\Fm$ is given by 
\begin{equation}
\Fm = \bfR^{\text{m}}\bfU^{\text{m}}=\bfV^{\text{m}}\bfR^{\text{m}}, 
\label{polardecomp}
\end{equation}
where $\bfR^{\text{m}}$ is a rotation, while $\bfU^{\text{m}}$  and $\bfV^{\text{m}}$ are
symmetric, positive-definite right and left stretch tensors. Consistent with standard notation, the total and the mechanical right Cauchy-Green deformation tensors are given by
\begin{equation}
\textbf{C} = \bfF^{\trans} \bfF \quad \text{and} \quad 
\Cm  = \Fmt \Fm. 
\label{cauchy_green}
\end{equation}
Using \eqref{kin1} and \eqref{mult_dec}, the velocity gradient may be written as
\begin{equation}
\bfL = \Lm + \Fm \Lc \Fmi
\label{velgrad1}
\end{equation}
with
\begin{equation}
\Lm = \Fmdot \Fmi, 
\qquad
\Lc = \Fcdot \Fci.
\label{velgrads2}
\end{equation}
We define the mechanical and chemical stretching and spin tensors through
\begin{equation}
\left.
\begin{split}
\Dm &= \Sym\Lm,\\[4pt]
\Dc &= \Sym\Lc,\\[4pt]
\end{split}
\qquad
\begin{split}
\bfW^{\text{m}} &= \skw\Lm,\\[4pt]
\bfW^{\text{c}} &= \skw\Lc,\\[4pt]
\end{split}
\right\}
\label{decomposition}
\end{equation}
such that $\Lm=\Dm+\bfW^{\text{m}}$ and $\Lc=\Dc+\bfW^{\text{c}}$.

Following Afshar and Di Leo \cite{afshar2021thermodynamically}, we employ a weighted decomposition of the chemical velocity gradient of the form 
\begin{equation}
    \Lc = \big(1-h(\xibar)\big)\Ld + h(\xibar) \Lg
    \label{vel_grad_dec}
\end{equation}
with $h(\xibar)$, an interpolation function of the reaction coordinate $\xibar$, which satisfies $h(0) = 0$ and 
$h(1) = 1$. This function is specified in Sect.~\ref{theory_spec} and consistently used for all quantities interpolated by the phase-field parameter $\xibar$.

The weighted decomposition \eqref{vel_grad_dec}, captures the combined deformation due to species transport across the solid host and distortion due to electrodeposition of a new material at the reaction sites. 
Here:
\begin{itemize}
    \item [i)] $\Ld$ captures the deformation of a material point due to transport of ionic species across the solid host;
    \item [ii)] $\Lg$ captures the deformation due to electrodeposition of a new material inside the solid host. 
\end{itemize}
Scaling by the interpolation function $h(\xibar)$ ensures that chemical distortions at material points in the bulk of the solid host, away form the reaction sites (i.e where $\xibar=0$) are only due to transport of ionic species, while in the fully-reacted state (i.e. where $\xibar=1$) due to electrodeposition of a new compound.

While the discussion on kinematics is so far presented in its most general form, we specialize our work specifically to modeling {\it single-ion conductors.} 
This class of  conductors is restricted to a single mobile cationic species transporting across the solid host, while anions remain immobilized in the backbone of the solid conductor.
In such a material, both {\it temporal and spatial concentration gradients are negligible within the bulk of the solid host} (i.e. away from the reaction sites), where deviations from electroneutrality (i.e. equilibrium bulk concentration) are insignificant \cite{dickinson2011electroneutrality,fuller2018electrochemical,bauer20113d}. 
Additionally, as detailed in  \cite{mcmeeking2019metal,klinsmann2019dendritic,barai2020role,deshpande2022models}, molar volume of Li-ions within the solid host electrolyte vanishes based on the notion that Li-ions lie within a rigid ceramic skeleton of the electrolyte that does not deform upon removal/addition of a Li atom. 
Henceforth, no deformations are associated with transport of ionic species across this class of solid ionic conductors, and hence $\Ld = \bf0$.

\vspace{10pt}
\begin{remark}
The choice to present the kinematic formulation in its most generic form is intended to showcase the versatility of the framework through the kinematic decomposition \eqref{vel_grad_dec} to model the phenomena of metal filament growth for the more general class of binary solid-ionic conductors (i.e. polymer solid electrolytes). For this class of solid conductors, chemical deformations may arise due to both a combination of species diffusion across the host and distortion due to electrodeposition of a new material at the reaction sites (c.f. Ganser et al. \cite{ganser2019finite}). 
\qed
\end{remark}

Focusing now on single-ion conductors, deformations associated with transport of ionic species across the solid conductor can be neglected, and \eqref{vel_grad_dec} simplifies to 
\begin{equation}
    \Lc = h(\xibar) \Lg.
    \label{vel_grad_dec2}
\end{equation}
Further, we make the pragmatic assumption that chemical deformation is irrotational (i.e. $\Wc = \bf0$) so that 
\begin{equation}
    \Dc = h(\xibar) \Dg,
    \label{dc_dg_rel}
\end{equation}
with $\Dg$, the electrodeposition-induced stretching assumed to depend on the reaction rate, $\xidotr$ through 
\begin{equation}
\Dg = \xidotr \Ng
\label{evoDg}
\end{equation} 
where $\Ng$ denotes the direction of electrodeposition-induced deformations.

Finally, the formulation presented so far makes no assumption regarding the nature of $\Fm$, which may be further decomposed to model, for example, the elastic-plastic behavior of the underlying solid conductor host and/or of the newly electrodeposited metal. 
We now {\it restrict our formulation such that all mechanical deformations are purely elastic in nature} and further specialize $\Fm = \Fe$, with $\Fe$ the elastic mechanical distortion. 
That is, both the solid conductor host and the newly electrodeposited metal will behave elastically. 

Finally, using \eqref{kin1}, \eqref{mult_dec}, \eqref{vel_grad_dec2} and \eqref{evoDg} we may then rewrite \eqref{velgrad1} for future use as
\begin{equation}
( \nabla \dot{\y} ) \bfF^{-1} =  \Fedot \Fei  
+ h(\xibar) \Fe  ( \xidotr \Ng)  \Fei.
\label{Defconstraint_rate}
\end{equation}

\vspace{10pt}
\begin{remark}
With regards to the solid ionic conductor, the assumption of purely elastic deformations is consistent with their stiff brittle nature.
On the other hand, continuous deposition of metallic material (e.g. Li-metal) inside defects induces large compressive stresses, which can cause the metal to flow plastically.  As the metal flows plastically, it may extrude from the crack to the electrode, limiting the pressure that can build up inside the crack and in turn the filament growth behavior.

As discussed by Klinsmann et al. \cite{,klinsmann2019dendritic} and Bucci and Christensen \cite{bucci2019modeling}, the possibility for the metal to plastically flow out of the crack mouth is impractical for realistic battery designs, suggesting that metal remains confined within the crack. 
In such a case, pressure will build up in the metal-filled defect until further damage occurs or the  reaction is suppressed by mechanical stresses, causing the metal deposition to cease. 
The nature and relevant importance of Li-metal plastic flow on filament growth through solid state electrolytes remains a topic of active research.    
In this work, we restrict our theoretical formulation to modeling Li-metal deformations as purely elastic. 
Owing to the already complex phenomena being modeled in this work, we purposely leave the addition of modeling an elastic-plastic metal phase to a future publication.  
\qed
\end{remark}

%%%%%%%%%%%%%%%%%%%%%%%%%%%%%%%%%%%%%%%%%%%%%%%%%%%%%%%%%%%%%%%%%%%%%%%%%%%
%%%%%%%%%%%%%%%%%%%%%%%%%%%%%%%%%%%%%%%%%%%%%%%%%%%%%%%%%%%%%%%%%%%%%%%%%%%
%%%%%%%%%%%%%%%%%%%%%%%%%%%%%%%%%%%%%%%%%%%%%%%%%%%%%%%%%%%%%%%%%%%%%%%%%%%
\section{Governing Balance Laws}
\label{sect:bal_laws}

In this section, we develop the governing equations for our theoretical framework, including macroscopic and microscopic force balances and thermodynamic laws.
For conciseness, we relegate portions of the detailed development to Appendix A, and present here the critical aspects of developing the governing balance laws. 

%%%
%%%
%%%
\subsection{Principle of virtual power. Balance of forces}
\label{sect:macro_micro_vir_pow}

To develop the remaining balance laws in our theoretical framework, we invoke the virtual-power approach (cf. Gurtin et al. \cite{gurtin2010mechanics}). This results in a macroforce balance and microforce balances for the rate-like kinematical descriptors in our theory. 
In exploiting the principle of virtual power, we consider a list of generalized virtual velocity fields to be given by
\begin{equation}
\calV = ( \delta \y, \delta \Fe,  \delta \xir, \nabla \delta \xir, \delta \dam, \nabla \delta \dam ).
\label{eq:Virt_Vel}
\end{equation} 
In light of  \eqref{Defconstraint_rate}, the virtual velocities are not independent and must obey the following kinematic constraint
\begin{equation}
( \nabla \delta \y ) \bfF^{-1}  = \delta \Fe \Fei +  h(\xibar)\Fe  ( \delta\xir \Ng) \Fei 
\label{Defconstraint_vir}
\end{equation}
%%%%%%%
For any part $\mathcal{P}$ of the reference body $\mathcal{B}$, the internal and external power are formulated in a consistent fashion with the manner in which forces expand power on $\mathcal{P}$,
%%%%%%%%%
\begin{equation}
\begin{split}
\delta W_\text{ext} (P,\calV) &= \int_{\partial\p} \bft_\mat(\bfn_\mat) \cdot \delta  \y  \, da_\mat 
%%%%
+  \int_{\p} \bfb_\mat \cdot \delta  \y \, dv_\mat 	
%%%%%
+ \int_{\partial \p} \eta \delta \xir \, da_\mat 
%%%%%
+ \int_{\partial \p} \gamma \delta \dam \, da_\mat \\[4pt]
%%%%%%%%%
%%%%%%%%%
\delta W_\text{int} (P,\calV) &= \int_{\p} (
\Se \tendot \delta \Fe
%%%%%%%%
+ E  \delta \xir	
%%%%%%
+\bfG \cdot \nabla \delta \xir 
%%%%%%%
+ \varpi  \delta \dam 
%%%%%%%%
+ \bfzeta \cdot \nabla \delta \dam ) \, dv_\mat 
\end{split}
\label{eq:VirtPower}
\end{equation}
%%%%%%%%
%%%%%%%%
where we have defined the following macroscopic and microscopic force systems conjugate to the kinematical variables,
%%%%%%
\begin{itemize}
\item[a)] A stress $\Se$ that expends power over the elastic distortion rate
$\dot{\bfF}^\text{e}$;
\item[b)] A traction $\bft_\mat(\bfn_\mat)$ (for each unit vector $\bfn_\mat$) that expends power over the velocity $\dot{\y}$;
\item[c)] A scalar microscopic stress $E$ that expends power over the rate $\xidotr$;
\item[d)] A vector microscopic stress $\bfG$ that expends power over the gradient $\nabla \xidotr$
\item[e)] A scalar microscopic traction $\mt(\bfn_\mat)$ that expends power over $\xidotr$ on the boundary of the part;
\item[f)] A scalar microscopic stress $\varpi$ that expands power over the rate $\damdot$;
\item[g)] A vector microscopic stress $\bfzeta$ that expands power over the gradient $\nabla\damdot$; and
\item[h)] A scalar microscopic traction $\gamma(\bfn_\mat)$ that expands power over $\damdot$ on the boundary of the part. 
\end{itemize}
Note that consistent with the discussion in Sect. \ref{sect:Kinematics}, no deformations are associated with the transport of ionic species in single-ion conductors. As a result, no mechanical power is associated with the concentration variable $\cc$. \newline

The principle of virtual power consists of two basic requirements: 
\begin{itemize}
%
%%%
\item[i)] Power balance, which requires that $ \delta W_\text{ext} (P,\calV) =  \delta W_\text{int} (P,\calV)$ for all generalized virtual velocities $\calV$
%
%%%
\item[ii)] Frame-indifference, which requires that $\delta W_\text{int} (P,\calV)$ is invariant under all changes in frame. 
\end{itemize}
We relegate a detailed development of the derivation of macroforce and microforce balances for the rate-like kinematic descriptors in our theory to Appendix A. Here, we succinctly summarize the  macroforce balance for the Piola stress, $\T_\mat$  as well as the two corresponding microforce balances for the stresses $\{E,\bfG, \bfzeta, \varpi \}$. These are given by
\begin{equation}
\left.
	\begin{split}
	&\text{Div} \, \T_\mat + \bfb_\mat = 0,
	\quad \text{with boundary condition}\quad 
	\bft_\mat(\bfn_\mat) = \T_\mat \bfn_\mat, \\[4pt]
	&E  -  \Jc h(\xibar) \Me \tendot \Ng  -  \text{Div} \bfG = 0,
	\quad \text{with boundary condition}
	\quad
	\eta(\bfn_\mat) = \bfG \cdot \bfn_\mat, \\[4pt]
	&\text{Div} \bfzeta - \varpi= 0,  
	\quad \text{with boundary condition}
	\quad
	\gamma(\bfn_\mat) = \bfzeta \cdot \bfn_\mat.
	\end{split}
	\, \right\}
	\label{eq:pvp_summ}
	\end{equation}
In writing \eqref{eq:pvp_summ}, we have introduced the classical Piola stress $\T_\mat = J \T \bfF^{-\trans}$, with $\T$ the Cauchy stress. In addition, we define the elastic Mandel stress as
\begin{equation}
    \Mme \Def \Je \Fet \T \Feit.
    \label{Mandel_stress}
\end{equation}
The macro and microforce balances, when supplemented with a set of thermodynamically-consistent constitutive equations, provide the governing mechanical equations for the theory. We note here that a discussion on invariance properties of the fields, consequence of the invariance of internal power under a change in frame, is omitted here for the sake of briefness. 
We refer the reader to the works of Anand \cite{anand2012cahn} and Afshar and Di Leo  \cite{afshar2021thermodynamically} for detailed derivations in the context of similar diffusion-reaction-deformation frameworks.

%%%
%%%
%%%
\subsection{Balance of energy. Entropy imbalance. Free energy imbalance}
\label{sect: bal_energ_entropy}

Our discussion of thermodynamics involves the following fields
\begin{center}
		\begin{tabbing}
			nnnnnnnnnnnnnnnnnn \= nnnn \= \kill
			\> $\ep_\mat$ \> the internal energy density per unit reference volume,\\[4pt]
			\> $\eta_\mat$ \> the entropy  density per unit reference volume,\\[4pt]
			\> ${\bfq_\mat}$ \> the heat flux   per unit reference area,\\[4pt]
			\> $q_\mat$ \> the external heat supply   per unit reference volume,\\[4pt]
			\> $\vartheta$ \> the absolute temperature $(\thet>0)$,\\[4pt]
			\> $\muM$ \> the chemical potential, \\[4pt]
			\> $\phi$ \>  the electric potential,
		\end{tabbing}
\end{center}
and follows the discussion of Gurtin et al. \cite{gurtin2010mechanics}. 
Consider a material region $\mathcal{P}$ and assume inertial effects (i.e. kinetic energy) to be negligible. Under {\it isothermal conditions}, the two laws of thermodynamics reduce to a single statement requiring that the  rate of increase in free energy of any part $\mathcal{P}$  is less than or equal to the power expended on $\mathcal{P}$.
Letting $\psi_\mat$  denote the free energy per unit reference volume, this requirement takes the form of a free-energy imbalance,
\begin{equation}
	\dot{\overline{\int_{\p}\!\psi_\mat\,dv_\mat}} \le  \calW_\text{ext}(\p)
   -\int_{\partial\p} \muM \bfj_\mat \cdot \bfn_\mat \, da_\mat
	-\int_{\partial\p} F\phi \bfj_\mat \cdot \bfn_\mat \, da_\mat - \int_{\partial\p} \phi \dmatdot \cdot \bfn_\mat \, da_\mat.
	\label{eng_imbalance1}
\end{equation}
Here, the second term on the right hand side of \eqref{eng_imbalance1} represents the flux of energy carried into $\mathcal{P}$ by the flux, $\bfj_\mat$ of the ionic species transporting across the solid host.
Additionally, following Kovetz \cite{kovetz2000electromagnetic}, the power expended on $\mathcal{P}$ by external charges is represented by an electromagnetic energy flux, which in the electrostatic limit, is given by the last two boundary integral terms on the right hand side of $\eqref{eng_imbalance1}$.

Consistent with electrochemical notation, we define 
\begin{equation}
\mu^{\text{e}} \Def \mu + F\phi
\label{electro_chem_pot}
\end{equation}
to denote the electrochemical potential of the mobile ionic species transporting across the solid host, with $F$ the Faraday constant. 
Making use of \eqref{electro_chem_pot}, one can alternatively write the energy imbalance \eqref{eng_imbalance1} as
\begin{equation}
	\dot{\overline{\int_{\p}\!\psi_\mat\,dv_\mat}} \le    \calW_\text{ext}(\p)
	-\int_{\partial\p} \mu^{\text{e}} \bfj_\mat \cdot \bfn_\mat  da_\mat 
    -\int_{\partial\p} \phi \dmatdot \cdot \bfn_\mat \, da_\mat.
	\label{eng_imbalance2}
\end{equation}
Defining the elastic second Piola stress as, 
\begin{equation}
    \Te \Def \Je \Fei \T \Feit =  \Cei \Me
	\label{second_Piola}
\end{equation}
with $\Ce = \Fet \Fe$ the elastic right Cauchy-Green tensor, the stress power $\Se\tendot\Fedot$ in \eqref{eq:VirtPower} may be written as $\Se\tendot\Fedot = (1/2) \Jc \Te \tendot \Cedot$. 
Equating the external power with the internal power \eqref{eq:VirtPower}, and applying the divergence theorem to the boundary integral terms, the energy imbalance \eqref{eng_imbalance2} may be written as
\begin{equation}
\begin{split}
\int_{\p} \left(
\dot{\psi}_\mat
- \frac{1}{2} \Jc \Te \tendot \Cedot 
- E \xidotr
- \bfG \cdot \nabla \xidotr 
- \varpi \damdot 
- \bfzeta \cdot \nabla\damdot 
+ \mu^{\text{e}} \Div \bfj_\mat   
+ \phi \Div \dmatdot 
+ \nabla \phi \cdot \dmatdot 
+  \bfj_\mat  \cdot \nabla \mu^{\text{e}}
\right) dv_\mat \le 0
\end{split}
\label{eng_imbalance3}
\end{equation}
Making use of mass balance \eqref{mass_bal1}, charge balance \eqref{char_balance}, Gauss' law \eqref{gauss_law}, along with the definition of electric field \eqref{Faraday_law2} and electrochemical potential \eqref{electro_chem_pot}, and using the fact that this inequality must hold for all parts $\mathcal{P}$, we obtain the local form of the free energy imbalance as
\begin{equation}
	\dot{\psi}_\mat  
	- \frac{1}{2} \Jc \Te \tendot \Cedot 
	- E \xidotr
	-  \bfG \cdot \nabla \xidotr -\varpi\damdot-\bfzeta\cdot\nabla\damdot
	- \mu \cdotr -  \mu^{\text{e}} \xidotr
	- \bfe_\mat \cdot \dmatdot
	+ \bfj_\mat \cdot \nabla \mu^{\text{e}}
	\le 0.
	\label{eng_imbalance5}
\end{equation}
For later use, we define the dissipation density $\calD \geq 0$ per unit volume per unit time as
\begin{equation}
	\calD
	= 
	\frac{1}{2} \Jc \Te \tendot \Cedot 
	+ E \xidotr 
	+  \bfG \cdot \nabla \xidotr +\varpi\damdot+\bfzeta\cdot\nabla\damdot
	+ \mu \cdotr + \mu^{\text{e}}  \xidotr
	+ \bfe_\mat \cdot \dmatdot
	- \bfj_\mat \cdot \nabla \mu^{\text{e}}
	- \dot{\psi}_\mat  
	\ge 0.
	\label{diss_ineq}
\end{equation}
Finally, we note here that all quantities in the free energy imbalance \eqref{eng_imbalance5} and dissipation inequality \eqref{diss_ineq} are invariant under a change in frame (c.f. Anand \cite{anand2012cahn} and Afshar and Di Leo \cite{afshar2021thermodynamically}).

%%%%%%%%%%%%%%%%%%%%%%%%%%%%%%%%%%%%%%%%%%%%%%%%%%%%%%%%%%%%%%%%%%%%%%%%%%%
%%%%%%%%%%%%%%%%%%%%%%%%%%%%%%%%%%%%%%%%%%%%%%%%%%%%%%%%%%%%%%%%%%%%%%%%%%%
%%%%%%%%%%%%%%%%%%%%%%%%%%%%%%%%%%%%%%%%%%%%%%%%%%%%%%%%%%%%%%%%%%%%%%%%%%%

\section{Constitutive Theory}
\label{Const_Theory}

We divide this section into energetic and dissipative constitutive equations.

%%%
%%%
%%%
\subsection{Energetic Constitutive Equations}
\label{energ_const_eqns}

Guided by the free-energy imbalance \eqref{eng_imbalance5}, we consider a set of constitutive equations for the free energy $\psi_\mat$, the stress $\Te$, the chemical potential $\mu$, the vector microforce $\bfzeta$ and the electric field $\bfe_\mat$ of the form
\begin{equation}
\psi_\mat =\hat{\psi}_\mat(\Lambda),
\quad 
\Te = \hat{\bfT}^{\text{e}}(\Lambda),
\quad
\mu = \hat{\mu}(\Lambda),
\quad
\bfzeta = \hat{\bfzeta}(\Lambda), 
\quad \text{and} \quad
\bfe_\mat = \hat{\bfe}_{\mat}(\Lambda),
\label{basicce}
\end{equation}
where $\Lambda$ denotes the list
\begin{equation}
\Lambda = (\Ce,\ccr , \xir,\nabla \xir, \dmat, \dam,  \nabla \dam).
\label{lambda_list}
\end{equation}
Substituting the constitutive equations \eqref{basicce} in the free-energy imbalance \eqref{eng_imbalance5} yields
\begin{equation}
	\begin{split}
	\left( \pards{\hat\psi_\mat}{\Ce} - \dfrac{1}{2} \Jc \Te \right) \tendot \Cedot
	+ \left(\pards{\hat{\psi}_\mat}{\ccr} - \muM \right) \cdotr
	+ \left(\pards{\hat{\psi}_\mat}{\xir} - E - \mu^{\text{e}} \right ) \xidotr
	+ \left(\pards{\hat{\psi}_\mat}{\nabla \xir} - \bfG \right) \cdot \nabla \xidotr \\
	+\left(\pards{\hat{\psi}_\mat}{\dam} - \stressen \right) \damdot 
	+ \left(\pards{\hat{\psi}_\mat}{\nabla \dam} - \bfzeta \right) \cdot \nabla \damdot 
	+ \left ( \pards{\hat{\psi}_\mat}{\dmat} - \bfe_\mat  \right ) \cdot \dmatdot
	- \stressdis \damdot
	+ \bfj_\mat \cdot \nabla \mu^{\text{e}} \le 0,
	\end{split}
	\label{thermo2a}
\end{equation}
where we have introduced a decomposition for the scalar damage microstress of the form
\begin{equation}
	\varpi = \stressdis + \stressen.
	\label{diss_additive}
\end{equation}
Motivated by  Anand \cite{anand2019modeling}, the decomposition \eqref{diss_additive} allows for both energetic and dissipative effects associated with temporal changes in $\dam$, while effects due to the gradient, $\nabla \dam$, a measure of the inhomogeneity of damage at the microscale, are taken to be entirely energetic.

From an electrochemical standpoint, in \eqref{thermo2a}, we make the distinction that processes associated with diffusion (governed by $\cdotr$) and migration (governed by $\dmatdot$) are energetic, while the ones associated with electrodeposition (governed by $\xidotr$) are dissipative. An exception to such choice is the power conjugate to $\nabla \xidotr$, which is taken to be entirely energetic (i.e all reaction dissipative processes are already accounted for in the term $\xidotr$).

As the inequality in \eqref{thermo2a} is to hold for all values of $\{\Cedot, \cdotr, \nabla\xidotr, \damdot, \nabla\damdot, \dmatdot \}$, their ``coefficients'' must vanish, for otherwise they may be chosen to violate \eqref{thermo2a}.
We are therefore led to the thermodynamic restriction that the free energy determines the stress $\Te$, the chemical potential $\mu$, the vector microstress $\bfG$, the scalar microstress $\stressen$, the vector microstress $\bfzeta$ and the electric field $\bfe_\mat$  through the ``state relations''
\begin{equation}
	\left.
	\begin{split}
	\Te&=  2 \Jci \pards{\hat{\psi}_\mat (\Lambda)}{\Ce}, \\
	\muM &= \pards{\hat{\psi}_\mat (\Lambda)}{\ccr},  \\
	\bfG &= \pards{\hat{\psi}_\mat(\Lambda)}{\nabla \xir},\\
	\stressen&= \pards{\hat{\psi}_\mat (\Lambda)}{\dam},\\
	\bfzeta &= \pards{\hat{\psi}_\mat(\Lambda)}{\nabla \dam}, \\
	\bfe_\mat& = \pards{\hat{\psi}_\mat(\Lambda)}{\dmat}. 
	\end{split}
	\right\}
	\label{energeticrelation}
\end{equation}
Recalling \eqref{electro_chem_pot} and using \eqref{energeticrelation}$_2$, the electrochemical potential of the ionic species transporting across the solid host, $\mu^{\text{e}}$ is given by 
\begin{equation}	
\mu^{\text{e}} = \pards{\hat{\psi}_\mat (\Lambda)}{\ccr} + F\phi.
\label{omega2}
\end{equation}

%%%
%%%
%%%
\subsection{Dissipative Constitutive Equations}	
\label{sect:diss_const}

The reduced dissipation inequality is now given by 
\begin{equation}
	\calD = 
	- \left( \pards{\hat{\psi}_\mat(\Lambda)}{\xir}  - E  -  \mu^{\text{e}} \right ) \xidotr
    + \stressdis \damdot
	- \bfj_\mat \cdot \nabla \mu^{\text{e}}   \ge 0.
	\label{dissipation}
	\end{equation}
We define the electrochemical potential of the electrodeposited species as
\begin{equation}
	\muxi  \Def  \pards{\hat{\psi}_\mat(\Lambda)}{\xir}  - E.    \label{elecdep_potential}
\end{equation}
Then, consistent with the notion of electrochemical reactions, we define the thermodynamic driving force for electrodeposition as a difference in electrochemical potential of the species participating in the reaction through 
\begin{equation}
	\mathcal{F} \Def  \muxi -  \mu^{\text{e}}. 
	\label{dep_driv_force}
\end{equation}
Using \eqref{dep_driv_force} in \eqref{dissipation}, the dissipation inequality may be rewritten as, 
\begin{equation}
	\calD = 
	- \calF \xidotr
    + \stressdis \damdot
	+ \nabla \mu^{\text{e}} \cdot \bfM_{\text{mob}}  \nabla \mu^{\text{e}}   \ge 0,
	\label{dissipation3}
\end{equation}
where guided by the dissipation inequality, we introduce a Fick-type relation for the flux, $\bfj_\mat$ of ionic species across the solid host of the form
\begin{equation}
	\bfj_\mat = - \bfM_{\text{mob}} \nabla \mu^{\text{e}}
	\label{flux1}
\end{equation}
where as is standard, the flux, $\bfj_\mat$ is linearly proportional to $\nabla \mu^{\text{e}}$ through the mobility tensor $\bfM_{\text{mob}} $.

Finally, we assume all terms in \eqref{dissipation3} to individually satisfy,
\begin{equation}
	\begin{split}
	- \calF \xidotr &\ge  0, \\
	\stressdis \damdot &\ge 0, \\
	\nabla \mu^{\text{e}} \cdot \bfM_{\text{mob}} \nabla \mu^{\text{e}}  &\ge 0.
	\end{split}
	\label{dissipation4}
\end{equation}
%%%%%%%
%
With respect to dissipation due to electrodeposition \eqref{dissipation4}$_1$, we assume, consistent with electro-kinetics theory,  that $\xidotr > 0$ if and only if $\calF < 0$, and vice versa $\xidotr < 0$ if and only if $\calF > 0$ (c.f. Bazant \cite{bazant2013theory}, Fuller et al. \cite{fuller2018electrochemical}). 
Given the constraint \eqref{dam_dot} for damage irreversibility  such that $\damdot \geq 0$, we satisfy \eqref{dissipation4}$_2$ by requiring $\stressdis \ge 0$. 
Finally, \eqref{dissipation4}$_3$ leads to the restriction that the mobility tensor $\bfM_{\text{mob}}$ is positive semi-definite. 
Altogether, these restrictions ensure that the dissipation inequality \eqref{dissipation3} is not violated.

%%%%%%%%%%%%%%%%%%%%%%%%%%%%%%%%%%%%%%%%%%%%%%%%%%%%%%%%%%%%%%%%%%%%%%%%%%%
%%%%%%%%%%%%%%%%%%%%%%%%%%%%%%%%%%%%%%%%%%%%%%%%%%%%%%%%%%%%%%%%%%%%%%%%%%%
%%%%%%%%%%%%%%%%%%%%%%%%%%%%%%%%%%%%%%%%%%%%%%%%%%%%%%%%%%%%%%%%%%%%%%%%%%%

\section{Summary of the general constitutive theory}
\label{sect:summ_general}

In this section, we summarize our general phase-field diffusion-reaction-damage  theory. 
The theory relates the following fields: 
\begin{center}
	\begin{longtable}{ll}
		$\bfx=\bfchi(\bfX,t) $, &     motion;\\[2pt]
		$\bfF=\nabla\bfchi, \quad J=\det\bfF>0$, &    deformation gradient;\\[2pt]
		$\bfF=\Fe \Fi$,   &   multiplicative decomposition of $\bfF$;\\[2pt]
		$\Fe, \quad J^e = \det\Fe > 0$, & elastic distortion; \\[2pt]
		$\Fi,  \quad \Ji=\det\Fi > 0$, &  chemical distortion;\\[2pt]
		$\bfL = \dot{\bfF}\bfF^{-1} = \Le + \Fe \Li \Fei$ & velocity gradient;\\[2pt]
		$\Li = \Fidot \Fii = \Di + \bfW^\text{c} \quad \text{with} \quad \bfW^\text{c} = 0$ & chemical velocity gradient;\\[2pt] 
		$\Di = h(\xibar) \Dg$ & chemical stretching; \\[2pt] 
		$\Dg = \xidotr \Ng$, & electrodeposition induced stretching; \\[2pt]
		$ \Fe =\bfR^{\text{e}}\bfU^{\text{e}}=\bfV^{\text{e}}\bfR^{\text{e}}$,    & polar decompositions of $\Fe$; \\[2pt]
		$\Ce = (\Fe)^{\trans}\Fe=(\bfU^{\text{e}})^2$, & right Cauchy-Green tensor;\\[2pt]
		$\bfT = \bfT^{\trans}$, & Cauchy stress; \\[2pt]
		$\Mme = \Je \Fet \T \Feit $, & elastic Mandel stress; \\[2pt]		
		$\bfT_\mat = J\bfT\F^{-\trans}$, & Piola  stress; \\[2pt]
		$\Te = \Je \Fei \T \Feit  $, & elastic second Piola stress; \\[2pt]
		$\psi_\mat $, & free energy density per unit reference volume; \\[2pt]
		$\cc$, & number of moles of diffusing ionic species per unit reference volume;\\[2pt]
		$\xi$, & number of moles of electrodeposited species per unit reference volume; \\[2pt]
		$\xibar = \xi / \ximax \in [0,1]$, & extent of the reaction; \\[2pt]
		$\nabla \xi$, & gradient of the reacted species concentration; \\[2pt]
		$\mue$, &  electrochemical potential of the diffusing species;\\[2pt]
		$\bfj_\mat$, & referential species flux vector. \\[2pt]
    $\bfe_\mat$, & referential electric field; \\[2pt]
    $\phi$, & electric potential; \\[2pt]
    $\bfd_\mat$ & referential electric displacement; \\[2pt]
    $\dam$ & phase-field damage variable; \\[2pt]
    $\nabla \dam$ & gradient of the damage variable.
	\end{longtable}
\end{center}

%%%
%%%
\subsection{Kinematics and Free Energy}

The deformation gradient is decomposed as 
\begin{equation}
\bfF = \Fe\Fi,
\label{}
\end{equation}
with $\Fe$ the elastic distortion, and $\Fc$ the chemical distortion. Further, the chemical velocity gradient is given by $\Li = \Fidot \Fii = \Di$, with the chemical stretching $\Di$ specialized as
\begin{equation}
\Di = h(\xibar) \Dg
\quad \text{and} \quad
\Dg = \xidot \Ng.
\label{}
\end{equation}
Here, $\Dg$ is the electrodeposition induced stretching, with $\xidot$ the reaction rate, $\Ng$ the direction of electrodeposition induced deformations, and $h(\xibar)$, an interpolation function, which satisfies $h(0) = 0$ and $ h(1) = 1$. 
The free energy is given by
\begin{equation}
\psi = \hat{\psi}(\Lambda),
\quad \text{with $\Lambda$, the list} \quad
\Lambda = (\Ce,\ccr , \xir,\nabla \xir, \dmat, \dam,  \nabla \dam).
\label{}
\end{equation}
%
%%%%%%%%%%%%%%%%
With the direction of electrodeposition induced deformations, $\Ng$ and the free energy function, $\hat{\psi}(\Lambda)$ prescribed, the following quantities may be derived.

%%%
%%%
\subsection{Stress. Force balance}
\label{sect:stress_summ}

The second Piola stress is given by \eqref{energeticrelation}$_1$
\begin{equation}
\Te =  2 \Jci \pards{\hat{\psi}_\mat (\Lambda)}{\Ce}
\label{eq:Te_summ}
\end{equation}
with the Cauchy stress given by $\bfT = \Jei \Fe \Te \Fet$, 
the elastic Mandel stress by $\Mme = \Je \Fet \T \Feit$,
and the first Piola stress by $\bfT_\mat = J \bfT \bfF^{-\trans}$.

The stress is governed by force balance \eqref{eq:pvp_summ}$_1$, viz. 
\begin{equation}
    \text{Div} \, \T_\mat + \bfb_\mat = 0,
    \label{eq:stress_summ}
\end{equation}
with Neumann boundary condition \eqref{eq:pvp_summ}$_1$, $\bft_\mat(\bfn_\mat) = \T_\mat \bfn_\mat$, where $\bft_\mat$ is a prescribed traction. 

%%%
%%%
\subsection{Electrochemical potential. Flux. Mass Balance}

Using \eqref{energeticrelation}$_2$ in \eqref{electro_chem_pot} and recalling \eqref{flux1}, the electrochemical potential and referential flux for the mobile ionic species are given by 
\begin{equation}
    \mu^{\text{e}} = \pards{\hat{\psi}_\mat (\Lambda)}{\ccr} + F\phi
    \quad \text{and} \quad
    \bfj_\mat = - \bfM_{\text{mob}} \nabla \mu^{\text{e}}
    \label{eq:mue_summ}
\end{equation}
with $\bfM_\text{mob}$, a positive  semi-definite mobility tenor. 

The concentration of ionic species transporting across the solid host, $c$ is governed by mass balance \eqref{mass_bal1}, viz.
\begin{equation}
    \cdotr = -\Div \bfj_\mat - \xidotr.
    \label{eq:mass_bal_summ}
\end{equation}
with $\xidot$, the species electrodeposition rate at the reaction site.

%%%
%%%
\subsection{Reaction driving force. Electrodeposition Kinetics}

Combining \eqref{elecdep_potential},\eqref{dep_driv_force} \eqref{eq:pvp_summ}$_2$ and \eqref{energeticrelation}$_3$ yields the electrodeposition driving force as 
\begin{equation}
    \calF = \muxi - \mue,
    \quad \text{with} \quad
    \muxi = \dfrac{\partial \hat{\psi}(\Lambda)}{\partial \xi} - \Ji h(\xibar)\Me\tendot\Ng - \Div \left(\dfrac{\partial \hat{\psi}(\Lambda)}{\partial \nabla \xi} \right).
    \label{eq:react_summ}
\end{equation}
The reaction kinetics are constrained by the dissipation inequality \eqref{dissipation4}$_1$ to obey $-\calF \xidot \geq 0$. Note that due to the phase-field (gradient) nature of the extent of reaction $\xibar$, \eqref{eq:react_summ} constitutes a PDE.
%
%%%
%%%
\subsection{Electrostatics}

The electric potential $\phi$ is governed by Guass' Law. Combining \eqref{gauss_law} with \eqref{char_balance} and  \eqref{energeticrelation}$_6$ with \eqref{Faraday_law2} yields
\begin{equation}
    \Div(\bfd_\mat) = F(\cc - \cco),
    \quad \text{with} \quad
    \bfe_\mat = \dfrac{\partial \hat{\psi}(\Lambda)}{\partial \bfd_\mat}, 
    \quad \text{and} \quad
    \bfe_\mat = - \nabla \phi(\bfX,t).
\label{eq:elect_summ}
\end{equation}
%%%
%%%
\subsection{Damage}

Combining \eqref{eq:pvp_summ}$_3$ with \eqref{energeticrelation}$_{4,5}$ and recalling the decomposition of the microstress $\varpi = \stressdis + \stressen$ from \eqref{diss_additive} yields the governing equation for the phase-field damage variable as,
\begin{equation}
    \Div\left( \dfrac{\partial \hat{\psi}(\Lambda)}{\partial \nabla \dam} \right) - \dfrac{\partial \hat{\psi}(\Lambda)}{\partial \dam} - \stressdis = 0,
    \label{eq:dam_summ}
\end{equation}
where the constraint $\stressdis \geq 0$ is required by the dissipation inequality \eqref{dissipation4}$_2$. 
%

%%%%%%%%%%%%%%%%%%%%%%%%%%%%%%%%%%%%%%%%%%%%%%%%%%%%%%%%%%%%%%%%%%%%%%%%%%%
%%%%%%%%%%%%%%%%%%%%%%%%%%%%%%%%%%%%%%%%%%%%%%%%%%%%%%%%%%%%%%%%%%%%%%%%%%%
%%%%%%%%%%%%%%%%%%%%%%%%%%%%%%%%%%%%%%%%%%%%%%%%%%%%%%%%%%%%%%%%%%%%%%%%%%%

\section{Specialization of the constitutive equations}
\label{theory_spec}

The theory developed above and summarized in Sect.~\ref{sect:summ_general} is quite general.
We introduce now a set of specialized constitutive equations to elucidate: 
i) the process of plating and fracture due to electrodeposition of Li-metal in single-ion conductors, specifically a LLZO inorganic electrolyte, and 
ii) the inherent coupling of electric, chemical and mechanical effects on electrodeposition kinetics and evolution morphology of Li-metal filaments across the solid-state electrolyte. 

The general electrodeposition reaction \eqref{Li_reac} is now specialized for Li-metal platting as
\begin{equation}
    \text{Li}^+ + e^- \rightarrow \text{Li}.
    \label{eq:Li_reac_spec}
\end{equation}

\subsection{Electrodeposition-Induced Deformation}
\label{elct_dep_def}

We begin by specifying the manner in which electrodeposition of Li-metal at the reaction sites induces mechanical deformations, specifically the direction tensor $\Ng$ in \eqref{evoDg}. 
Consistent with Narayan and Anand \cite{narayan2020modeling} and Rejovitzky et. al \cite{rejovitzky2015theory}, we consider a generic distortion due to electrodeposition such that
\begin{equation}
    \Fgdot = \Dg\Fg.
\end{equation}
Additionally, we assume the electrodeposition induced stretching tensor $\Dg$ is anisotropic and specialize it as
\begin{equation}
	\Dg = \epsgdot \Sg \quad \text{where} \quad \Sg \Def \mg \otimes \mg.
	\label{anisotropy_dg}
\end{equation}
Here, the unit vector, $\mg$  represents the direction of electrodeposition induced deformations,
while $ \epsgdot = \text{tr}(\Dg) $ denotes the total volumetric strain rate for reaction induced deformations.
Consistent with experimental observations \cite{ganser2019extended}, we assume deformations due to electrodeposition of Li-metal to occur preferentially in a direction normal to the reaction front (e.g. normal to the conductor/metal interface), and hence specialize $\mg$ as 
\begin{equation}
    \mg = \nabla \xibar / |\nabla \xibar|.
\end{equation}

Further, we assume  volumetric distortions due to electrodeposition to vary linearly with the extent of reaction (c.f \cite{narayan2021coupled,anand2012cahn}), such that $\Jg = 1+\Omega\xir$. 
Then, using the fact that $\dot{\Jg}\Jgi = \Tr(\Dg)$, and combining with \eqref{anisotropy_dg} yields 
\begin{equation}
	\Dg = \frac{\Omega\xidotr}{1+\Omega\xir}\mg\otimes\mg.
	\label{eq:anis_strech}
\end{equation}
Here, $\Omega$ denotes the constant partial molar volume associated with formation of Li-metal due to electrodeposition and may be calculated from experiments or ab-initio simulations. 
Given \eqref{eq:anis_strech} and recalling that $\Dg = \xidot \Ng$ from \eqref{evoDg}, the direction of electrodeposition induced deformations is specialized as
\begin{equation}
	\Ng =\frac{\Omega}{1 + \Omega\xir } \mg\otimes\mg.
\label{N_growth}
\end{equation}

\subsection{Free Energy}
\label{free_enrg}
Next, we consider a separable free energy per unit volume of the form
\begin{equation}
\hat{\psi}_\mat(\Ce,\ccr,\xir,\nabla \xir,\dmat,\dam,\nabla\dam) 
 = \psimech(\Ce,\xir,\dam)  
 + \psic(\ccr)  
 + \psi_\mat^\xi(\xir) 
 + \psi_\mat^{\nabla\xi}(|\nabla\xir|)
 + \psi_\mat^e(\Ce,\xi,\dmat)
 + \psi_\mat^\dam(|\nabla \dam|) 
\label{free_energ_decom}
\end{equation}
All individual functions in \eqref{free_energ_decom} will be specialized as isotropic functions of their arguments. 
Here:
\begin{itemize}
%
%%% Mechanical Freen Enegy
%%%
\item[(i)] 
$\psimech(\Ce,\xi,\dam)$  is the contribution to changes in the free energy density due to elastic deformation of the host  material. 
A consequence of isotropy is that the free energy $\psimech$ may be expressed as (cf. Anand \cite{anand2012cahn}),
\begin{equation}
    \psimech(\Ce,\xi,\dam) = \psimech(E_1^\text{e},E_2^\text{e},E_3^{\text{e}},\xi,\dam),
\end{equation}
where $\{E_1^\text{e},E_2^\text{e},E_3^\text{e}\}$ denote the principal values of the logarithmic elastic strain defined as
\begin{equation}
	\Ee = \text{ln} \Ue = \sum_{i=1}^{3} \Eei \bfr_i^{\text{e}} \otimes \bfr_i^{\text{e}} = \sum_{i=1}^{3} \text{ln} \lambdaei \; \bfr_i^{\text{e}} \otimes \bfr_i^{\text{e}}.
	\label{princ_dec_E}
\end{equation}
Further, given \eqref{princ_dec_E} above, straightforward calculations (c.f. Anand and Su \cite{anand2005theory}) show that the mechanical contribution to Mandel stress\footnote{We refer to this as the ``mechanical" contribution to the Mandel stress as it is based only on the derivative of the mechanical free energy, $\psi_\mat^\text{m}$, and not the entire free energy, $\psi_\mat$. As such, it omits certain possible stress contributions, such as the Maxwell stress due to electrostatics. A detailed discussion on the significance of these contributions is presented in Remark 4 below.} is then given by 
\begin{equation}
	\Mme =  \Jci \sum_{i=1}^{3} \pards{{\psi}_\mat^\text{m} (E_1^\text{e},E_2^\text{e},E_3^\text{e}, \xir, \dam) }{\Eei} \bfr_i^{\text{e}} \otimes \bfr_i^{\text{e}}
    \label{Mandel_iso}
\end{equation}
We further decompose our free energy formulation due to elastic deformations into a ``positive" part, $\psimechpos$ due to tension and a  ``negative" part, $\psimechneg$ due to compression (cf. Miehe et al.\cite{miehe2010phase,}, Klinsmann et al. \cite{klinsmann2016modeling}, and Navidtehrani et al. \cite{navidtehrani2021simple}) this yields
\begin{equation}
\psimech(\Ee,\xir,\dam) = g(\dam)\psimechpos(\Ee,\xir) + \psimechneg(\Ee,\xir)
\label{energ_split}
\end{equation}
where $g(\dam)$ is a degradation function of the damage variable, $\dam$.

The "positive" and "negative" mechanical parts of the free energy are individually specialized to the classical strain energy function of isotropic linear elasticity to moderately large elastic deformations using the logarithmic strain measure, $\Ee$ (c.f. Anand \cite{anand1979h}).
They are given by
\begin{equation}	
\left.
\begin{split}
&\psimechpos(\Ee,\xir) \Def \Jc 
\left[ 
G(\xibar) \bigg( \langle  \Eeone \rangle _{+}^{2} + \langle  \Eetwo \rangle _{+}^{2} +
\langle  \Eethree \rangle _{+}^{2} \bigg) 
+ \frac{1}{2} \left( K(\xibar) - \frac{2}{3} G(\xibar) \right) \bigg( \langle  \Eeone +  \Eetwo  + \Eethree \rangle_{+} \bigg)^{2} 
\right] \\[4pt]
& \psimechneg(\Ee,\xir) \Def \Jc 
\left[ 
G(\xibar) \bigg( \langle  \Eeone \rangle _{-}^{2} + \langle  \Eetwo \rangle _{-}^{2} +
\langle  \Eethree \rangle _{-}^{2} \bigg) 
+ \frac{1}{2} \left( K(\xibar) - \frac{2}{3} G(\xibar) \right) \bigg( \langle  \Eeone +  \Eetwo  + \Eethree \rangle_{-} \bigg)^{2} 
\right]
\end{split}
\right\}
\label{eq:psi_split}
\end{equation}
where, $G(\bar\xi)$ and $K(\bar\xi)$ denote the reaction-dependent shear and bulk moduli respectively.
Consistent with \eqref{princ_dec_E}, $\{\Eei| i = 1,2,3\}$ denote the principal elastic logarithmic strains. For notational convenience in \eqref{eq:psi_split}, we additionally introduce
\begin{equation}
\langle \Eei \rangle_{+} =
		\begin{cases}
		\Eei & \text{if } \Eei > 0\\
		0                  & \text{otherwise}
		\end{cases}
		\quad \text{and} \quad
		\langle \Eei \rangle_{-} =
		\begin{cases}
		\Eei & \text{if } \Eei < 0\\
		0                   & \text{otherwise}
		\end{cases}
\end{equation}

We note here that only the ``positive" part of the free energy is degraded in \eqref{energ_split}, while the ``negative" part of the free energy remains undegraded.
This in turn prevents damage from occurring under purely compressive stresses.
The split in free energy due to elastic deformations is necessary not just to maintain resistance of the material under compression, but to additionally enable damaged regions, which have become electroplated (i.e. regions with $\dam=1$ and $\xibar=1$) to develop and sustain compressive stresses.

The degradation function is taken to be a monotonically decreasing function of $\dam$, which concurrently satisfies $g(0) = 1$ and $g(1) = 0$. Consistent with Miehe et al. \cite{miehe2010phase}, we adopt 
\begin{equation}
    g(\dam) = (1-\dam)^{2} + \epsilon,
\label{deg_function}
\end{equation}
where $\epsilon \approx 0$ is a small positive-valued constant introduced to prevent ill-conditioning of the model when $\dam = 1$. 
Finally, the reaction-dependent shear and bulk moduli $G(\bar\xi)$ and $K(\bar\xi)$ obey
\begin{equation}
\begin{split}
G(\xibar) &= \big(1-h(\xibar)\big) G^{\text{SE}} + h(\xibar) G^{\text{M}}, \\[2pt]
K(\xibar) &= \big(1-h(\xibar)\big) K^{\text{SE}} + h(\xibar) K^{\text{M}},
\end{split}
\label{g_k_variation}
\end{equation}
which, consistent with \eqref{vel_grad_dec}, are interpolated using the function $h(\xibar)$.
The specific form of $h(\xibar)$ is given by
\begin{equation}
    h(\xibar) = \xibar^3(6\xibar^2-15\xibar+10).
    \label{eq:hfunc}
\end{equation}
This form of the interpolation function \eqref{eq:hfunc} is widely used in the phase-field literature (c.f. Chen et al. \cite{chen2015modulation} and Guyer et al. \cite{guyer2004phase}).
Throughout, we invoke the ``SE" and ``M" superscripts to denote quantities associated with the solid-state electrolyte and the electrodeposited Li-metal respectively. 
%
%%%
%%% Chemical Free Energy
%%%
\item[(ii)] 
$\psic(\ccr)$ is the change in chemical free energy due to mixing of ionic species with the solid host. As a simple continuum approximation to mixing, we take the chemical free energy to be given by a regular solution model (cf. DeHoff \cite{dehoff2006thermodynamics}) as
\begin{equation}
\psic(\ccr)  = \muo \ccr  + R\,\thet\, \ccrmax \Big( \cbar \ln \cbar + (1 - \cbar) \ln(1-\cbar)\Big) 
\label{energymixing}
\end{equation}
Here, $\cbar = \ccr$/$\ccrmax$ with  $\cbar \in [0,1]$ represents the normalized species concentration in the solid host, while $\ccrmax$ denotes the maximum species concentration in moles per unit reference volume when all the accommodating sites in the host material are filled. 
Additionally, $\muo$ denotes the reference chemical potential of the diffusing species, $R$ the universal gas constant, and $\vartheta$, the absolute temperature.
%
%%% Reaction Free Energy
%%%
\item[(iii)] $\psi_\mat^\xi(\xi)$ includes  a double-well energy function associated with the energetic barrier across the phases in addition to an electrochemical energetic contribution associated with the standard (reference) chemical state and electrostatic potential of the electrodeposited Li-metal. 
It is given by
\begin{equation}
\psi_\mat^\xi (\xir) = \mathcal{W} \xibar^2 (1 - \xibar)^2
+  \xir \muxiz + \xir F \phi_{0}.
\label{energy_Lithium}
\end{equation}
Consistent with the works of Guyer et al.~\cite{guyer2004phase} and Cogswell \cite{cogswell2015quantitative}, the first term governed by $\calW$, sets the height of the energy barrier across the phases in our continuum kinetics formulation.
The second term defines the reference chemical potential, $\muxiz$ of the electrodeposited species. 
Finally, the third term represents the energetic contribution associated with the electrostatic potential of the metal, $\phi_0$ (cf. Chen et al. \cite{chen2015modulation} and Bucci et al. \cite{bucci2016formulation}).
%
%%% Inteface Energy 
%%%
\item[(iv)] 
$\psi_\mat^{\nabla\xi}(|\nabla \xi|)$ is the interfacial free energy, which penalizes sharp interfaces and is simply taken as
\begin{equation}
\psi_\mat^{\nabla\xi}(|\nabla\xi|)=  \frac{1}{2}\lambda_\xi |\nabla \xir|^2,
\label{energygradient}
\end{equation}
with $\lambda_\xi \geq 0$, a gradient energy coefficient. 
This energetic contribution regularizes the problem from a numerical standpoint by defining a finite interface width --- controlled by $\lambda_\xi$ ---  for the phase-field reaction coordinate, $\xibar$.
%
%%% Electrostatics
%%%
\item[(v)] $\psi_\mat^e(\Ce,\xi,\bfd_\mat)$ is the electrostatic energy, which is taken as (cf. Narayan et al.\cite{narayan2021coupled} and Suo \cite{suo2010theory})
\begin{equation}
    \psi_\mat^{e} = J \left( \frac{1}{2\varepsilon} \textbf{d} \cdot \textbf{d} \right).
    \label{spat_elec_energ}
\end{equation}
Here, $\varepsilon(\xibar)$ denotes the reaction-dependent effective electrical permittivity, and $\bfd$ is the electric displacement in the current configuration. This function is {\it isotropic in the deformed body} as it depends only on the magnitude of $\bfd$. 

Using $ \bfd= J^{-1} \bfF \bfd_\mat$, we may express \eqref{spat_elec_energ} in terms of the referential electric displacement as 
\begin{equation}
    \psi_\mat^{e}(\Ce,\dmat, \xir) = J^{-1} \left( \frac{1}{2\varepsilon} \dmat \cdot \textbf{C} \dmat \right)
    \label{electrostatic_energ}
\end{equation}

The reaction-dependent effective electrical permittivity is expressed as, 
\begin{equation}
\varepsilon(\xibar) = \varepsilon_{r}(\xibar)\varepsilon_{0}
\label{permittivity}
\end{equation}
where $\varepsilon_{0} = 8.85\cdot10^{-12}$ F/m denotes the electrical permittivity of vacuum and $\varepsilon_{r}(\xibar)$ denotes the reaction-dependent relative electrical permittivity of each material phase.
Consistent with other interpolations in this theory, we assume
\begin{equation}
    \varepsilon_{r}(\xibar)= \Big(1-h(\xibar)\Big)\varepsilon^{\text{SE}}_{r} +h(\xibar) \varepsilon^{\text{M}}_{r}.
\end{equation}
%
%%% Damage 
%%%
\item[(vi)] 
$\psi_\mat^{\dam}(|\nabla \dam|)$ is the free energy contribution associated with gradient effects of damage. 
We specialize its form as a quadratic function of the magnitude of the gradients in damage, $|\nabla \dam|$ such that
\begin{equation}
    \psi_\mat^\text{d}(|\nabla \dam|) = \frac{1}{2}\psicr \ell^2 |\nabla \dam|^2.
\label{grad_dam_energ}
\end{equation}
Here, $\psicr > 0$ is a coefficient with units of energy per cubic volume, while $\ell>0$ denotes an internal length scale, which sets the width of  damage zone across which the damage field  varies in a diffuse fashion.
%%%%%%%%%%%%%%%
\end{itemize}

Finally, combining \eqref{energymixing} through \eqref{grad_dam_energ}, the total free energy per unit volume accounting for the combined effects of mixing, reaction, electrostatics, finite elastic deformations and damage is given by
\begin{equation}
\begin{split}
{\psi}_\mat(\Ee,\ccr,\xir,\nabla \xir,\dmat,\dam,\nabla\dam) 
&= 
g(\dam)\psimechpos(\Ee,\xir) + \psimechneg(\Ee,\xir) \\ 
&+ \muo \ccr  + R\,\thet\, \ccrmax \left( \cbar \ln \cbar + (1 - \cbar) \ln(1-\cbar)\right) \\
&+ \mathcal{W} \xibar^2 (1 - \xibar)^2 
+  \xir \muxiz + \xir F \phi_{0} 
+  \frac{1}{2}\lambda_\xi |\nabla \xir|^2 \\
& + J^{-1}\left( \frac{1}{2\varepsilon} \dmat \cdot \textbf{C}  \dmat \right)
+ \frac{1}{2}\psicr \ell^2 |\nabla \dam|^2
\end{split}
\label{summ_free_energ}
\end{equation}
with $\psimechpos(\Ee,\xir)$ and $\psimechneg(\Ee,\xir)$ defined in \eqref{eq:psi_split}.
%%%
%%%
\subsection{Stress}
\label{stress_spec}
%%%%%%%%%%%%%%%%%

Using \eqref{summ_free_energ} in \eqref{Mandel_iso}, the Mandel stress is given as
\begin{equation}
\Mme = g(\dam)\Mmepos + \Mmeneg
\label{eq:Me_sep}
\end{equation}
with
\begin{equation}
\begin{split}
\Mmepos &= \sum_{i=1}^{3} 
\bigg[ 
2G(\xibar) \langle \text{E}_{\text{i}}^{\text{e}} \rangle_{+} 
+ \left( K(\xibar)-\frac{2}{3}G(\xibar) \right) \bigg( \langle  \Eeone +  \Eetwo  + \Eethree \rangle_{+} \bigg) 
\bigg] 
\bfr_i^{\text{e}} \otimes \bfr_i^{\text{e}} \\[2pt]
\Mmeneg &= \sum_{i=1}^{3} \Big [2G(\xibar) \langle \text{E}_{\text{i}}^{\text{e}} \rangle_{-} + \Big(K(\xibar)-\frac{2}{3}G(\xibar) \Big ) \Big( \langle  \Eeone +  \Eetwo  + \Eethree \rangle_{-} \Big) \Big ] \bfr_i^{\text{e}} \otimes \bfr_i^{\text{e}} 
\label{eq:Me_pos_neg}
\end{split}
\end{equation}
and the Cauchy and Piola stress measures related by
\begin{equation}
\begin{split}
\bfT = \Jei \bfRe \Mme \bfRet
\quad \text{and} \quad
\bfT_\mat &= \Jc \Big( \bfRe \Mme \bfRet \Big)\bfF^{-\trans}.
\end{split}
\label{eq:T_spec}
\end{equation}

\vspace{10pt}
\begin{remark}
We note here that in deriving the stress relations \eqref{eq:Me_sep} and \eqref{eq:Me_pos_neg}, we omit the contribution to stress due to electrostatics (i.e. the Maxwell stress). 
That is, we neglect the derivative of the elecrostatic energy $\psi_\mat^e(\Ce,\xi,\bfd_\mat)$ with respect to the elastic Cauchy-Green tensor. 
A detailed derivation of the Maxwell stress due to electrostatics, leading to the classical form  $\bfT^{\text{es}} = \varepsilon (\textbf{e}  \otimes \textbf{e} - (1/2) (\textbf{e}\cdot{\textbf{e})\mathbf{1}})$, with $\textbf{e}$ the spatial electric field, may be found in the work of Narayan et al. \cite{narayan2021coupled}. 
For the case of inorganic solid electrolytes, as modeled here, a relative permittivity value $\varepsilon_r \approx 50$ has been experimentally reported \cite{rettenwander2015synthesis}. 
In light of \eqref{permittivity}, the permittivity of the solid host evaluates then to a negligibly small number.
As detailed in the works of Natsiavas et al. \cite{natsiavas2016effect}, Shishvan et al. \cite{shishvan2020growth} and Narayan et al. \cite{narayan2021coupled}, the Maxwell stresses due to electrostatics are negligibly small compared to the stresses generated by the deposition of metal filaments inside the solid host and can be neglected without compromising accuracy. 
As such, we omit the contribution to mechanics due to electrostatics so as not to obfuscate the more relevant physics discussed next.
\qed
\end{remark}

%%%
%%%
\subsection{Electrochemical Potential. Flux}
\label{elec_chem_pot}

Using \eqref{summ_free_energ} in \eqref{eq:mue_summ}$_1$, the electrochemical potential of the ionic species transporting across the solid host is given by
\begin{equation}
    \mu^{\text{e}} = \muo + R\vartheta \text{ln} \left( \frac{\cbar}{1- \cbar} \right) + F\phi.
    \label{mu_species_sse}
\end{equation}

We specialize the flux of ionic species across the solid host to be isotropic and write \eqref{eq:mue_summ}$_2$ as 
\begin{equation}
    \bfj_\mat = - m(\ccr,\xibar)\nabla \mu^{\text{e}},
    \label{flux_spec}
\end{equation}
with the effective mobility, $m$ given by
\begin{equation}
{m}(\cc,\xibar) = {m}_{0}(\xibar)\ccr(1-\cbar), 
\quad \text{and} \quad 
{m}_0(\xibar) = \Big(1-h(\xibar)\Big) {m}_{0}^\text{SE} + h(\xibar){m}_{0}^\text{M}.
\label{eff_mobility}
\end{equation}
Here, ${m}_{0}^\text{SE}$ and ${m}_{0}^\text{M}$ are related to the diffusivities of each material phase through the standard relations,
\begin{equation}
   {m}_{0}^\text{SE} = \frac{D_{0}^\text{SE}}{R\vartheta}, 
   \quad \text{and} \quad  
   {m}_{0}^\text{M} = \frac{D_{0}^\text{M}}{R\vartheta}
\end{equation}
%%%
with $D_{0}^\text{SE} $ and $D_{0}^\text{M}$ denoting the diffusivity of ionic species in the solid host and the electrodeposited Li-metal.

%%%
%%%
\subsection{Reaction Driving Force. Electrodeposition Kinetics}
\label{reac_spec}

Using \eqref{summ_free_energ} in \eqref{eq:react_summ}$_2$, the electrochemical potential of the deposited Li-metal is given by
\begin{equation}
\begin{split}
\muxi &=  \muxiz
+ \dfrac{d}{d\xir} \bigg( \mathcal{W} \xibar^2 (1-\xibar)^2 \bigg)  
+  F \phi_{0} 
- J^c h(\xibar)\Me \tendot \Ng - \Div(\lambda_{\xi} \nabla \xi).
\label{muxi_elec_dep}
\end{split}
\end{equation}
We note that in deriving \eqref{muxi_elec_dep}, we neglect the derivatives of $J^c$ and the reaction-dependent shear, $G(\xi)$ and bulk modulus $K(\xi)$, present in the mechanical free energy, $\psi_\mat^{\text{m}}$, with respect to the reaction coordinate, $\xi$. These terms are quadratic in the elastic strains, and expected to be much smaller in magnitude than the other terms in \eqref{muxi_elec_dep} (c.f. Narayan and Anand \cite{narayan2022coupled}, Di Leo et al. \cite{di2015diffusion}).

Further, using \eqref{mu_species_sse} and \eqref{muxi_elec_dep} in \eqref{eq:react_summ}$_1$, the thermodynamic driving force $\calF$ for electrodeposition takes the following form
\begin{equation}
\begin{split}
\calF &= \muxi- \mu^{\text{e}} \\
&= 
\underbrace{ 
(\muxiz - \muo)
}_{\text{energetic}} 
- \underbrace{
R\vartheta \left( \frac {\cbar}{1-\cbar} \right) 
}_{\text{entropic}}
+ \underbrace{
F (\phi_0 - \phi)
}_{\text{electric}} 
+  \underbrace{
\dfrac{d}{d\xir} \bigg( \calW \xibar^2 (1-\xibar)^2 \bigg) 
} _{\text{energetic barrier}} 
- \underbrace{ 
\Ji h(\xibar) \Me \tendot \Ng
}_{\text{mechanical}}
- \underbrace{
\Div(\lambda_{\xi} \nabla \xir) 
}_{\substack{\text{numerical} 
\\[1pt] \text{regularization}}}.
\end{split}
\label{reac-evo}
\end{equation}
Here, the first term denotes the difference in reference chemical potentials between species transporting across the solid host and species in the electrodeposited metallic compound. 
The second term captures the role of configurational entropy. 
Critically, enrichment in concentration of ionic species at the reaction sites favors electrodeposition of Li-metal, while  depletion in concentration of Li-ions accordingly retards electrodeposition. 
The third term serves as an electric driving force for electrodeposition, stemming from the difference in electric potential in the electrode and electric potential in the solid conductor at the reaction site.
The fourth term introduces a local driving force to the reaction associated with the energy barrier between the phases, which drives the reaction towards the two energy minima.

The fifth term captures the effect of mechanical stress on the reaction driving force. In light of the discussion in Sect. \ref{elct_dep_def}, only the stress component normal to the electrode-solid conductor interface affects the driving force for electrodeposition (c.f. Ganser et al. \cite{ganser2019extended} and Deshpande and McMeeking \cite{deshpande2022models}). This coupling is different in form to conventional diffusion-deformation theories employing an isotropic chemical distortion, in which case the reaction driving force couples to mechanics through a pressure term (c.f. \cite{di2014cahn,di2015diffusion}). 
Critically, compressive stresses retard the electrodeposition of metallic compound at the reaction sites, while tensile stresses accordingly facilitate plating. The final term in \eqref{reac-evo} arises from the gradient phase-field nature of the theory and regularizes the interface width, setting a minimum width controlled by $\lambda_\xi$.  
%%%%%%%%%%

Having discussed the reaction driving force, $\calF$, we now specify the reaction kinetics equation. Consistent with electro-kinetics theory, we invoke a Butler-Volmer non-linear reaction kinetics formulation and evolve the extent of electrodeposition, $\xibar$ as follows,
\begin{equation}
	\dot{\xibar} = 
	\begin{cases}
	R_0 \left( \exp \left( \dfrac{-\alpha\calF}{R\vartheta} \right) 
	- \exp\left( \dfrac{(1-\alpha)\calF}{R\vartheta} \right) \right), 
	& \quad \text{if} \quad 0 < \xibar < 1, \\[2pt]
	0, & \quad \text{if} \quad \xibar = 1.
	\end{cases}
	\label{xi_evolution}
\end{equation}
%%%
Here, $\alpha$ denotes a symmetry factor, representative of the fraction of surface overpotential promoting anodic or cathodic reaction at the electrode interface, while $R_0 > 0$ denotes a positive-valued reaction constant. 
For elementary single-electron transfer reactions, a value $\alpha = 0.5$ is typically employed.
Note that \eqref{xi_evolution} satisfies the dissipation inequality \eqref{dissipation4}$_1$, such that $-\calF \xidot \geq 0$, where electrodeposition of Li metal, $\dot{\xibar} > 0 $ proceeds under a negative reaction driving force, $\calF < 0$, consistent also with electro-kinetics theory. 

We emphasize that the focus of this work is on modeling growth of metal filaments across a solid conductor with continuous deposition (i.e. $\dot{\xibar} > 0$). 
As a result, \textit{we do not model the reverse “stripping” process (i.e. $\xibardot < 0$) and assume perfect contact between the electrode and the solid conductor is maintained at all times.}

%%%
%%%
\subsubsection{Restrictions on electrodeposition kinetics}

We now introduce two phenomenological restrictions on the electrodeposition kinetics formulation presented above. 
First, the phase-field reaction formulation presented here does not explicitly account for the presence of electrons in the electrodeposition reaction \eqref{eq:Li_reac_spec}, which are necessary for Li-metal plating to occur. 
Since the solid-state electrolyte is electronically-insulating, we assume that electrons are readily available only at the Li metal/SSE interface. 
We account for this phenomenon by restricting the electrodeposition reaction \eqref{xi_evolution} to occur {\bf only} in regions where there is existing Li-metal deposits (and henceforth readily available electrons), modeled as regions with $\xibar > 0$. 
Phenomenologically, we model this behavior by scaling the reaction rate constant, $R_0$ with a reaction-dependent logistic function, $f_1(\xibar)$ of the form

\begin{equation}
    f_1(\xibar) = g_1(\xibar) - g_1(0),
    \quad \text{with} \quad
    g_1(\xibar) = \left( 1+\text{exp}\Big( -\alpha_1(\xibar-\beta_1)\Big) \right)^{-1}.
    \label{coupling_electron}
\end{equation}
The logistic function \eqref{coupling_electron} is guaranteed to satisfy $f_1(0) = 0$, such that no Li-metal platting will occur in the solid-state electrolyte bulk where no electrons are present.
The parameters $\alpha_1$ and $\beta_1$ in \eqref{coupling_electron} respectively control the steepness of the logistic function and the midpoint location.

Second, as discussed in Sect.~\ref{sect:Bal_Eqns}, we model platting to occur only within damaged regions of the solid state electrolyte, which provide the necessary vacant sites to accommodate Li-metal deposition.
The theoretical formulation thus far does not guarantee that $\xidot > 0$ only in regions where $\dam > 0$. 
To phenomenologically enforce this constraint, we once again scale the reaction constant, $R_0$ by a damage-dependent logistic function of the form
\begin{equation}
    f_2(\dam) = g_2(\dam) - g_2(0),
    \quad \text{with} \quad
    g_2(\dam) = \left( 1+\text{exp}\Big( -\alpha_2(\dam-\beta_2)\Big) \right)^{-1}.
    \label{coupling_damage}
\end{equation}
Scaling of the reaction kinetics by the logistic function $f_2(\dam)$ shown in \eqref{coupling_damage} ensures that electrodeposition of Li-metal inside the solid conductor is confined only within regions of the host electrolyte which have incurred some degree of damage. 

Given these constraints, the reaction kinetics formulation \eqref{xi_evolution} is replaced with the modified form  
\begin{equation}
	\dot{\xibar} = 
	\begin{cases}
	f_1(\xibar)f_2(\dam)R_0\left( \exp \left( \dfrac{-\alpha\calF}{R\vartheta} \right) 
	- \exp\left( \dfrac{(1-\alpha)\calF}{R\vartheta} \right) \right), 
	& \quad \text{if} \quad 0 < \xibar < 1, \\[2pt]
	0, & \quad \text{if} \quad \xibar = 1,
	\end{cases}
	\label{xi_evolution_mod}
\end{equation}
with the logistic functions defined in \eqref{coupling_electron} and \eqref{coupling_damage}.

\begin{figure}[h!]
    \centering    \includegraphics[width=5.3in]{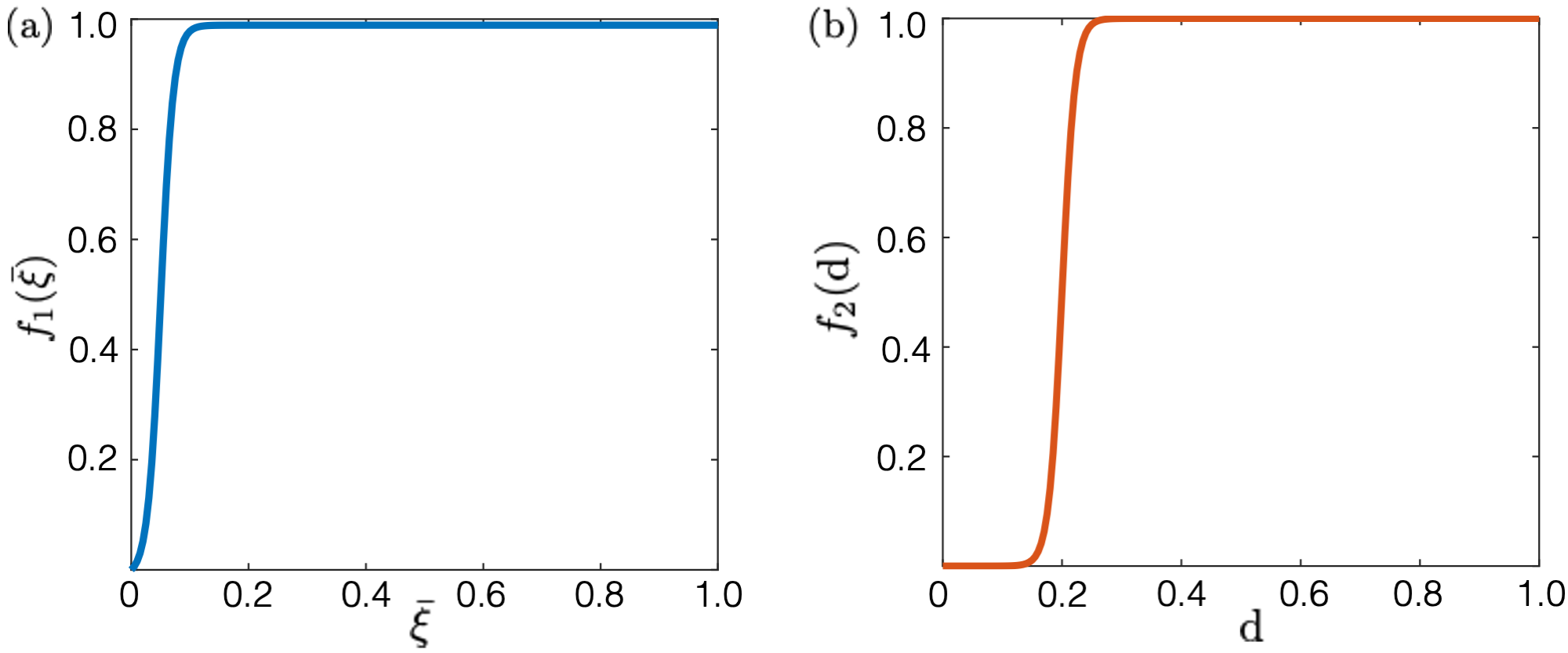}
    \caption{Logistic functions modulating the reaction constant $R_0$ with (a) the function $f_1(\xibar)$ and (b) the function $f_2(\dam)$.}
    \label{fig:rate_modifiers}
\end{figure}

The specific logistic functions chosen are shown in Fig.~\ref{fig:rate_modifiers}.
For both functions, we choose $\alpha_1 = \alpha_2 = 90$ to ensure a steep transition. 
For the function $f_1(\xibar)$, which phenomenologically models the requirement of electrons being present, we choose a small value of the offset point $\beta_1 = 0.05$. 
This suffices to enforce that Li-metal plating occurs in the presence of existing Li-metal deposits (i.e. in the presence of electrons), without altering the reaction kinetics significantly.
For the function $f_2(\dam)$, which restricts electrodeposition within damaged regions of the host electrolyte, we specialize $\beta_2 = 0.2$.
This essentially requires that $\approx$ 20\% damage is incurred on the solid electrolyte before Li-metal platting can occur.
In Sect. \ref{num_sims} below, we provide a detailed discussion on the role of the logistic the function, $f_2(\dam)$ which couples reaction kinetics and damage. 

%%%
%%%
\subsection{Electrostatics}
\label{elec_field_spec}

Using \eqref{summ_free_energ} in \eqref{eq:elect_summ}$_2$, the referential electric field is given by
\begin{equation}
    \emat = \frac{1}{\varepsilon}J^{-1} \textbf{C} \dmat.
    \label{elc_fld_ref}
\end{equation}
%%%%%%
Invoking the relations $\textbf{e} = \bfF^{-\trans} \emat$ and $\textbf{d} = J ^{-1} \bfF \dmat$, one may also relate the spatial electric flux density to the spatial electric field as 
\begin{equation}
    \textbf{d} = \varepsilon \textbf{e}.
    \label{spat_elct_disp}
\end{equation}

%%%
%%%
\subsection{Damage}
\label{dam_evol}

The evolution equation for damage is adopted from the works of Anand et al. \cite{anand2019modeling} and Miehe et al. \cite{miehe2010phase}. 
For brevity, we omit detailed derivations and refer the reader to the aforementioned publications.
We note that using \eqref{summ_free_energ} in \eqref{eq:dam_summ} and defining the dissipative microstress $\stressdis$ as
\begin{equation}
    \stressdis = \psicr + \Gamma \damdot,
\end{equation}
the final evolution equation for damage takes the form 
\begin{equation}
\Gamma\damdot = 2(1-\dam) \mathcal{H}-\psicr\dam 
+ \psicr \ell^2 \Delta\dam. 
\label{dam_pde}
\end{equation}
Here, $\Gamma >0 $ denotes a small viscous regularization parameter, introduced to impart stability to the numerical solution scheme. 
Consistent with the derivations in Anand et al. \cite{anand2019modeling}, in the rate-independent limit ($\Gamma = 0$),  the energy dissipated per unit volume as damage, $\dam$ increases from 0 to 1 evaluates to $\psicr$. 
Naturally then, in our formulation, $\psicr$ represents an energy per unit volume dissipated during the fracture process. 
As later discussed in Sect. \ref{num_sims}, $\psicr$ can be related to the experimentally reported fracture energy of the material of choice, $G_{c}$ through a characteristic length scale parameter, $\ell$, which controls the diffuse fracture zone.

Finally, $\mathcal{H}$ represents a monotonically increasing history field function, which ensures that the damage irreversibility constraint \eqref{dam_dot} holds. 
It is given by
\begin{equation}
    \mathcal{H}  =  \max\limits_{\text{s} \in \left[0, \text{t}\right]} \left[\langle{\psimechpos(\Ee,\xi)   - \psicr/{2}}\rangle \right].
	\label{history_func}
\end{equation}
Consistent with \eqref{energ_split}, only the mechanical energetic contribution associated with tensile strains, $\psimechpos$, contributes to the evolution of damage.

%%%%%%%%%%%%%%%%%%%%%%%%%%%%%%%%%%%%%%%%%%%
%%%%%%%%%%%%%%%%%%%%%%%%%%%%%%%%%%%%%%%%%%%
%%%%%%%%%%%%%%%%%%%%%%%%%%%%%%%%%%%%%%%%%%%

\section{Governing partial differential equations for the specialized constitutive equations.  Boundary conditions} 
\label{sect:PDE-sec}

The final set of governing partial differential equations consists of:
\begin{enumerate}
%
%%%
\item
Force balance, eq. \eqref{eq:stress_summ},  viz.
\begin{equation}
\Div \T_\mat + \bfb_{\mat} = \zed, 
\label{PDE_mech}
\end{equation}
with the Piola stress $\T_\mat$ given by \eqref{eq:T_spec}$_2$, and $\bfb_{\mat}$ the non-inertial body force.
%
%%%
\item
The local mass balance for the ionic species \eqref{eq:mass_bal_summ}, which together with the flux relationship \eqref{flux_spec} yields
\begin{equation}
\cdotr = \Div(m \nabla \mu^{\text{e}} ) - \xidotr,
\label{mass_bal_pde}
\end{equation}
with the mobility, $m$, given in \eqref{eff_mobility}, and the electrochemical potential of ionic species, $\mu^{\text{e}}$, in \eqref{mu_species_sse}. 
The reaction rate, $\xidot$, is governed by the PDE \eqref{pde_reaction} below.

%
%%%
\item 
The reaction kinetics, governed by \eqref{xi_evolution_mod}, which in conjunction with the reaction driving force \eqref{reac-evo}, yield the governing PDE for the reaction coordinate
\begin{equation}
\begin{split}
\dot{\xibar} &= f_1(\xibar)f_2(\dam)R_0 \left(\exp\left(\dfrac{-\alpha\calF}{R\vartheta}\right)  - \exp\left(\dfrac{(1-\alpha)\calF}{R\vartheta}\right)\right), \quad \text{with} \\[2pt]
\calF &= 
\begin{split}  
(\muxiz - \muo)
- 
R\vartheta\left(\frac {\cbar}{1-\cbar}\right) 
+ 
F (\phi_0 - \phi)
+
\dfrac{d}{d\xir} \Big( \mathcal{W} \xibar^2 (1-\xibar)^2 \Big) 
-
\Ji h(\xibar) \Me \tendot \Ng
-
\Div(\lambda_{\xi} \nabla \xir).
\end{split}
\label{pde_reaction}
\end{split}
\end{equation}
Here, eq. \eqref{pde_reaction} constitutes a PDE due to the presence of the $\Div (\nabla \xir)$ term in the reaction driving force  $\calF$.
Following \eqref{xi_evolution_mod}, the reaction kinetics also obey the restriction $\dot{\xibar} = 0$  when $\xibar = 1$, which marks the end of the reaction.
%
%%%
\item
Gauss' law, viz. \eqref{eq:elect_summ}, which along with electric displacement given by \eqref{elc_fld_ref}, and using $\bfe_\mat = -\nabla \phi$ governs the electric potential, $\phi$ through
\begin{equation}
\Div(\dmat) = F(\ccr - \ccro), 
\quad \text{with} \quad
\dmat = -\varepsilon J \textbf{C}^{-1} \nabla \phi.
\label{elec_pot_pde}
\end{equation}

From a computational standpoint, numerically solving for $\phi$ using \eqref{elec_pot_pde} is cumbersome and requires resolution of thin boundary layers in the range of a few nanometers, where deviations from electroneutrality occur \cite{fuller2018electrochemical,dickinson2011electroneutrality}.
Given the representative dimensions of the domains of interest in this work, this makes the problem computationally non-tractable and expensive. 

In practice, granted the regions where concentration gradients occur are small with respect to the domain of interest, an assumption of electroneutrality in the system is often invoked in the literature to numerically solve for $\phi$, such that the net charge $\qr = F(\ccr - \ccro) = 0$ (cf. \cite{chen2015modulation, narayan2022coupled,hong2018phase,tantratian2021unraveling, yuan2021coupled}).

We adopt this approach in our numerical implementation.
Making use of \eqref{charge_balance_v2}, we consider the current density, $\imat$ to be conserved and described by a Poisson equation including a sink term to represent the charge loss due to electrochemical reactions (c.f. \cite{chen2015modulation,tantratian2021unraveling,hong2018phase}), yielding a governing equation for $\phi$ of the form,
\begin{equation}
\Div\imat = - F\xidotr 
\quad \text{with} \quad 
\imat = -\kappa(\xibar) \nabla \phi
% \quad \text{and} \quad 
\label{pde_elec_pot2}
\end{equation}
which replaces the use of \eqref{elec_pot_pde}.
Here, consistent with electro-kinetics theory, we introduce the effective conductivity, $\kappa(\xibar)$, which interpolates between the conductivity of the electrode, $\kappa^\text{SE}$, and the conductivity of the Li metal, $\kappa^\text{M}$, through $\kappa(\xibar) = (1-h(\xibar))\kappa^{SE} + h(\xibar)\kappa^{M}$.

%
%%%
\item

Damage evolution, governed by eq. \eqref{dam_pde}, viz. 
\begin{equation}
	    \Gamma\damdot = 2(1-\dam) \mathcal{H}-\psicr\dam +  \psicr \ell^2 \Delta\dam 
\label{pde_damage}
\end{equation}
with $\mathcal{H}$ a monotonically increasing history function which ensures damage irreversibility (i.e. $\damdot \ge 0)  $ defined in \eqref{history_func}. 
\end{enumerate}

Finally, to complete the model, a set of boundary and initial conditions must be supplemented. 
Let $\calS_1$ and $\calS_2$ be complementary subsurfaces on the boundary $\partial\B$ of the body $\B$, i.e. $\partial \B=\calS_{\bfchi}\cup\calS_{\tmat}$ and $\calS_{\bfchi} \cap\calS_{\tmat}=\emptyset$.
Similarly, let the pairs $\{\calS_3,\calS_4\}$,  $\{\calS_5,\calS_6\}$, $\{\calS_7,\calS_8\}$, and $\{\calS_9,\calS_{10}\}$ be complementary subsurfaces as described above.
For the five degrees of freedom $\{\bfu, \cbar, \phi, \xibar, \dam\}$ governed by the PDEs described above, we then respectively consider the boundary conditions
\begin{equation}
\left.
\begin{split}
\bfu &=\breve{\bfu} \quad \text{on} \quad \calS_{1},
\quad \text{and} \quad 
\T_\mat\bfn_\mat = \breve{\bft}_\mat \quad \text{on} \quad \calS_{2}; \\
\cbar &= \breve{\cbar} \quad \text{on} \quad \calS_{3},
\quad \text{and} \quad
-\bfj_\mat \cdot \bfn_\mat =\breve{j} \quad \text{on} \quad \calS_{4};
\\
\phi &= \breve{\phi} \quad \text{on} \quad \calS_{5} ,
\quad \text{and} \quad
-\dmat \cdot \bfn_\mat =\breve{\varpi} \quad \text{on} \quad \calS_{6};
\\
\xibar & =\breve{\xibar} \quad \text{on} \quad \calS_{7},  
\quad \text{and} \quad
\lambda_{\xi} \nabla \xibar  \cdot \bfn_\mat = 0  \quad \text{on} \quad  \calS_{8};
\\
\dam &= \breve{\dam} \quad \text{on} \quad \calS_{9},
\quad \text{and} \quad
{\nabla\dam} \cdot \bfn_\mat = 0  \quad \text{on} \quad \calS_{10}.
\end{split}
\right \}
\label{eq:bcs}
\end{equation}
Here: i) $\breve{\bfu}$ is a prescribed displacement and $\breve{\bft}_\mat$ a prescribed traction, ii) $\breve{\cbar}$ is a prescribed normalized concentration and $\breve{j}$ a prescribed flux, iii) $\breve{\phi}$ is a prescribed electric potential and $\breve{\varpi}$ a prescribed surface charge density, iv) $\breve{\xibar}$ is a prescribed reaction coordinate, and v) $\breve{\dam}$ is a prescribed state of damage. 
Consistently, the initial conditions are also prescribed as
\begin{equation}
 \bfu(\X,0) = \bfu_0(\X), \,\, 
 \cbar(\X,0) = \cbar_0(\X), \,\, 
 \phi(\X,0) = \phi_0(\X), \,\,
 \xibar(\X,0) = \xibar_0(\X), \,\, \text{and} \,\,
 \dam(\X,0) = \dam_0(\X).
\label{init_cond}
\end{equation}
The fully coupled set of PDEs \eqref{PDE_mech} through \eqref{pde_damage}, along with the boundary conditions \eqref{eq:bcs} and initial conditions \eqref{init_cond} give then an initial/boundary-value problem for the unknowns of displacement $\bfu(\X, t)$, normalized concentration $\cbar(\X,t)$, electric potential $\phi(\X,t)$, reaction coordinate $\xibar(\X,t)$ and state of damage $\dam(\X,t)$.

%%%%%%%%%%%%%%%%%%%%%%%%%%%%%%%%%%%%%%%%%%%
%%%%%%%%%%%%%%%%%%%%%%%%%%%%%%%%%%%%%%%%%%%
%%%%%%%%%%%%%%%%%%%%%%%%%%%%%%%%%%%%%%%%%%%

\section{Numerical simulations}
\label{num_sims}

In this Section, we detail a set of numerical simulations aimed to highlight the important features of our theory, namely the interplay of electro-chemo-mechanical processes that govern the initiation and propagation of Li-metal filaments in SSEs. 
 While the proposed framework is general in nature, we specialize it towards modeling a problem of engineering relevance to commercialization of next-generation all-solid-state batteries as detailed below. 
 
The theoretical framework specialized in Sect.~\ref{theory_spec} contains five degrees of freedom $\{ \xibar,\cbar,\phi, \dam, \textbf{u} \}$ governed by the partial differential equations summarized in Sect.~\ref{sect:PDE-sec}.
The fully-coupled PDEs are solved using the finite element method in Abaqus Standard \cite{systemes2010abaqus} by developing a custom user-element (UEL) subroutine following the details described in Chester et al. \cite{chester2015finite}.
We note that implementation of the reaction kinetics PDE \eqref{pde_reaction} is done through the so called ``micromorphic" formulation, where an auxiliary variable is introduced to ease numerical convergence \cite{forest2016nonlinear,di2015chemo}.
The implementation is akin to that of Afshar and Di Leo \cite{afshar2021thermodynamically} and we refer the reader to this work for further details.
Importantly, we note that use of this numerical technique {\it does not affect} the numerical results, and produces equivalent outputs as would be generated with a direct implementation of \eqref{pde_reaction}.

We specialize our theory for a Li-metal/LLZO solid-state architecture, which has been widely investigated in the literature (c.f \cite{fincher2022controlling,porz2017mechanism,swamy2018lithium,ren2015direct,cheng2017intergranular}).
This architecture, using an oxide-based electrolyte, is relevant due to its superior chemo-mechanical properties (i.e.  high stiffness and fracture toughness, high ionic conductivity, excellent stability against Li-metal etc.).
Growth of Li-metal filaments across a LLZO solid-state electrolyte has been experimentally observed to occur as both a single transgranular branch, as well as through a network of filaments proceeding across the grain boundaries of the SSE \cite{swamy2018lithium,ren2015direct,chen2015modulation,heo2021short,porz2017mechanism}.

In Sect.~\ref{transg_growth}, we focus on modeling growth of a single Li-metal filament traversing from the anode towards the cathode end. 
For the set of simulations in Sect.~\ref{transg_growth}, we do not consider the LLZO microstructure composition (i.e. the presence of grain boundaries) and treat the solid host as a homogenized medium.
Starting with a small initial imperfection at the electrode interface, we demonstrate the manner in which the coupled electro-chemo-mechanical phenomena govern the growth of Li-metal filaments with continuous fracture of the SSE.
Additionally, we elucidate the role that mechanical confinement (i.e. mechanical boundary conditions) plays on the rate of crack propagation versus the rate of Li-metal filament growth.
We demonstrate the ability of our theory to qualitatively reproduce the experimentally observed phenomenon of crack fronts propagating ahead of Li-metal filaments, with cracks only partially filled by the electrodeposited Li-metal \cite{hao2021tracking,ning2021visualizing}.

In Sects.~\ref{dam_mech_kinetics}, we consider the role of the SSE microstructure composition by modeling the presence of grain boundaries through seeding of elements with varying properties. 
At present, there is a lack of numerical frameworks, which can incorporate the manner in which microstructural information affects the coupled electro-chemo-mechanical growth of Li-metal filaments through SSEs. 
We use this set of simulations to elucidate the manner in which damage and mechanical stresses are coupled to electrodeposition kinetics in this theoretical framework. In particular, we investigate how damage can impact electrodeposition through both the thermodynamically consistent reaction driving force $\calF$ in \eqref{reac-evo}, and the reaction kinetics restrictions in \eqref{xi_evolution_mod}. 
Finally, we demonstrate the utility of our theoretical framework by modeling the manner in which the morphology of Li-metal filaments changes for a LLZO microstructure depending on the relative fracture strength of the LLZO grains versus the grain boundaries.

%%%
%%%
%%%
\subsection{Modeling growth of a single Li-metal filament across a LLZO electrolyte. Role of mechanical confinement}
\label{transg_growth}

We present here an application of the fully-coupled, electro-chemo-mechanical framework towards modeling the growth of Li-metal filaments across a Li$_7$La$_3$Zr$_2$O$_{12}$ (LLZO) electrolyte. 
Growth of Li-metal filaments across pre-existing defects at the metal electrode interface for single-crystal LLZO microstructures has been extensively reported in the literature (c.f. \cite{porz2017mechanism,swamy2018lithium}). 
We demonstrate the ability of our framework to numerically reproduce the phenomena of Li-metal filament growth as a single branch protruding from the anode towards the cathode, while fracturing the SSE in the process. 
Through these simulations, we elucidate the interplay of electro-chemo-mechanical processes, which govern electrodeposition of Li-metal across SSEs.
In particular, we elucidate the manner in which mechanical confinement of the simulation domain plays a critical role on the resulting fracture and electrodeposition morphology.

\begin{figure}[h!]
    \centering
    \includegraphics[width=6.5in]{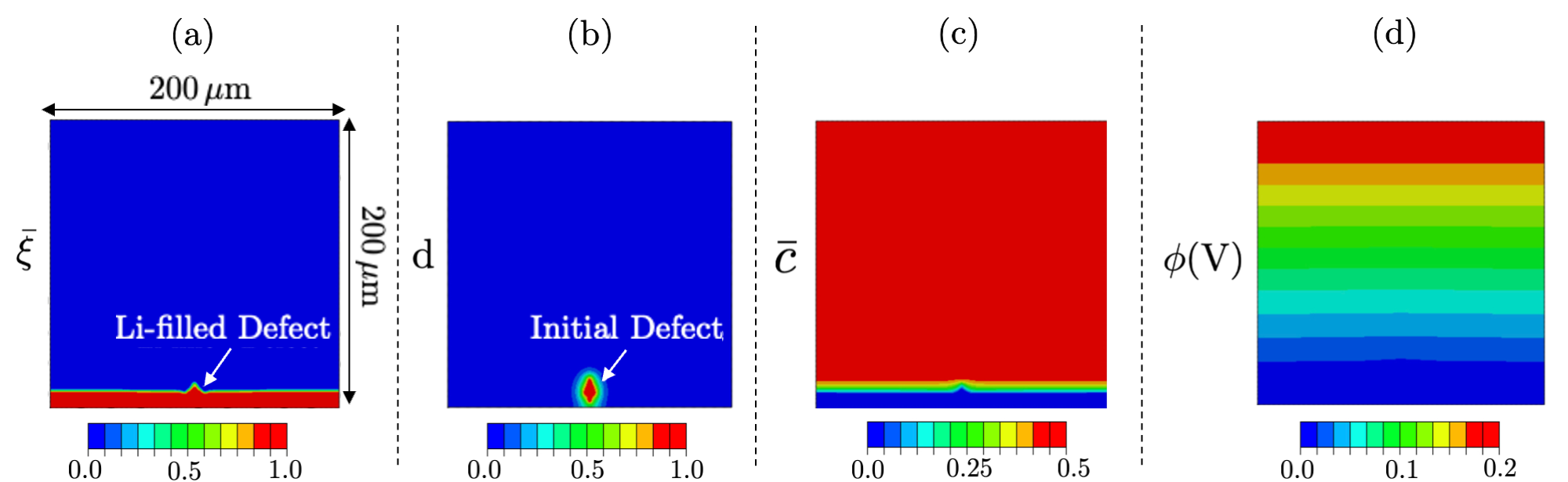}
    \caption{Initial configuration of the Li-electrode/solid electrolyte simulation containing a pre-existing defect. Contours of (a) reaction coordinate $\xibar$, (b) damage $\dam$, (c) normalized concentration $\cbar$, and (d) electric potential $\phi\,\text{(V)}$.}
   \label{fig:init_config}
\end{figure}

We consider a two-dimensional (2D), 200 $\mu$m x 200 $\mu$m plane-strain simulation domain, meshed with 350 x 350 quadrilateral finite elements.
This domain is representative of a half-cell system consisting of a Li-metal anode and a LLZO solid electrolyte, see Fig.~\ref{fig:init_config}(a). 
The dimensions of the simulation domain are assigned consistent with works in the literature invoking a similar phase-field treatment \cite{hong2018phase,tantratian2021unraveling,tian2019interfacial}. 
To simulate the Li-metal anode, a 5 $\mu$m thick layer (c.f. \cite{tantratian2021unraveling,narayan2020modeling}) is initialized in an already fully-reacted state at the bottom of the domain, Fig. \ref{fig:init_config}(a). 
We do so by prescribing a reaction coordinate,  $\xibar$ = 1 across this area.
To simulate the presence of a Li-filled crack-like defect on the electrolyte's surface\cite{porz2017mechanism,bucci2019modeling,klinsmann2019dendritic}, an imperfection 8 $\mu$m in width and 2 $\mu$m in height is seeded on the anode/SSE interface by prescribing a state of damage, $\dam = 1$ and reaction coordinate, $\xibar$= 1, see Fig.~\ref{fig:init_config}(b).
The prescribed defect dimensions are comparable to experimental works investigating the role of surface defects on metal growth across inorganic SSEs \cite{porz2017mechanism,wolfenstine2018mechanical}. 
This setup, Figs.~\ref{fig:init_config}(a,b), mimics infiltration of Li-metal in a pre-existing crack. 
As shown later, the Li-filled defect serves to break the symmetry of the model and initiate subsequent growth of Li-metal filaments across the SSE.

A vanishing normalized cation concentration, $\cbar \approx 0$ is prescribed across the anode layer, Fig. \ref{fig:init_config}(c), to numerically simulate a state of complete ionic species depletion in the pure Li-metal phase. 
According to the calculations of Li and Monroe \cite{li2019dendrite}, there are 15 available sites for lithium per La$_3$Zr$_2$O$_{12}$ formula unit, 7 of which are occupied at equilibrium, yielding an equilibrium occupancy  $\theta = x / x_{\text{max} }\approx 0.5$ for Li$_7$La$_3$Zr$_2$O$_{12}$.
Consistently, we prescribe a normalized bulk cationic concentration, $\cbar = 0.5$ as an initial condition across the solid electrolyte (regions of $\xibar = 0$), see Fig.~\ref{fig:init_config}(c).
The same normalized concentration, $\cbar = 0.5$, is additionally prescribed as a Dirichlet boundary condition on the top edge of the simulation domain. 
As shown in Fig. \ref{fig:init_config}(d), we set the anode potential at the bottom of the domain, $\phi = 0$V and prescribe a potential, $\phi = 0.20\,\text{V}$ on the top of the doman, consistent with works in the literature \cite{tian2019interfacial,ren2020inhibit,shen2021does}. 
The prescribed electric potential on the top edge of the domain, i.e. the electrolyte surface, serves to drive the system out of equilibrium, enabling for electrodeposition of Li-metal at the electrode-SSE interface. 
Altogether, Fig. \ref{fig:init_config}(a-d) constitute the initial, \textit{stress-free} configuration of the Li-metal electrode-solid electrolyte assembly.

Mechanically, roller boundary conditions are applied at the top edge of the simulation domain, while the bottom edge is pinned.
We consider two different boundary conditions for the left and right edges of the simulation domain to illustrate the role of mechanical confinement on electrodeposition kinetics and resulting fracture/electrodeposition morphology. 
It has been experimentally reported via in-situ X-ray tomography the propensity for Li-metal filaments to sometimes only partially fill cracked regions of SSEs (c.f. \cite{hao2021tracking,ning2021visualizing}). 
The cited experimental works have demonstrated the potential for electrodeposition-induced cracks to transverse the entire SSE ahead of the Li-deposits, which trail behind. 
To numerically reproduce this phenomenon, we apply first a roller boundary condition to the sides. 
Subsequently, we relax this lateral mechanical constraint and consider the sides to be free.

\vspace{10pt}
\begin{remark}
Note that the region being modeled as the Li-metal anode shown in Figs.~\ref{fig:init_config}(a,b) is characterized by a state of $\xibar = 1$ and $\dam = 0$, unlike the lithium-filled defect, which is characterized by $\xibar = 1$ and $\dam = 1$. 
As noted in Sect.~\ref{sect:Bal_Eqns}, in the development of this theoretical framework, we restrict Li-filament growth inside the SSE to occur only in damaged regions (i.e. $\xibar > 0$ only if $\dam > 0 $). 
Since the Li-metal anode is initialized before the start of the physical simulation (i.e. in a stress-free configuration), we do not require that $\dam > 0$ in this region. 
The Li-metal anode region is characterized by $\xibar = 1$ and $\dam = 0$ so as to model a solid Li-metal material capable of sustaining both tensile and compressive loads as would be physically present in a Li-metal/LLZO half cell. 
In addition, we note here that we allow for electrodeposition of Li-metal at the flat horizontal anode/SSE interface, which as discussed above are characterized by a state of $\xibar=1,\dam=0$. 
As such, in those regions {\it we omit} the logistic function $f_2(\dam)$ in the reaction kinetics \eqref{xi_evolution_mod} such that electroplating on the flat horizontal interfaces may occur. 
\qed
\end{remark}

The electro-chemo-mechanical properties for a Li-Metal/LLZO solid-state system are tabulated in Tab.~\ref{tab:props}. 
\begin{table}[h!]
\caption{Material properties for modeling of metal filament growth in a Li-Li$_7$La$_3$Zr$_2$O$_{12}$ system.}
\label{tab:props}
%
% \vspace{-0.2in}
%
% \begin{center}
\begin{tabular}{ p{1.2in} p{1.3in} p{1.6in} p{1.7in} }
\hline
& Parameter & Value & Source \Trule\Brule \\
\hline
Electro-Chemical & $\kappa_{\text{LLZO}}$ & 4.43$\cdot$10$^{-2}$ S/m & \citeauthor{li2019enhanced} \cite{li2019enhanced} \Trule\Brule\\
&$\kappa_{\text{Li}}$ & 1.0$\cdot$10$^{7}$ S/m & \citeauthor{chen2015modulation} (\cite{chen2015modulation}) \Trule\Brule\\
& $D_\text{LLZO}$ & 1.0$\cdot 10^{-12}$ m$^2$/s & \citeauthor{tian2019interfacial} (\cite{tian2019interfacial}) \Trule\Brule\\
& $D_\text{Li}$ & 1.0$\cdot 10^{-15}$ m$^2$/s & \citeauthor{tian2019interfacial} (\cite{tian2019interfacial}) \Trule\Brule\\
& c$_\text{max}$ & 4.22$\cdot 10^{4}$ mol/m$^3$ & \citeauthor{tian2019interfacial} (\cite{tian2019interfacial}) \Trule\Brule\\
%%%%%
%%%%%
Reaction Kinetics & $R_{0}$ & 0.1\, s$^{-1}$ & \citeauthor{yuan2021coupled} (\cite{yuan2021coupled}) \Trule\Brule\\
& $\Omega_\text{Li}$ & 1.3$\cdot 10^{-5}$ m$^3$/mol & \citeauthor{tantratian2021unraveling} (\cite{tantratian2021unraveling})
\Trule\Brule\\
& $\Omega_\text{Li} \ximax$ & 0.3 & This Work
\Trule\Brule\\
& $\mathcal{W}$ & 1.18$\cdot10^6$\, J/m$^3$ & \citeauthor{tian2019interfacial} (\cite{tian2019interfacial}) \Trule\Brule\\
& $\lambda_{\xi}$ & 8$\cdot10^4$\, pJ$\cdot$$\mu$m$^5$/pmol$^2$ & \text{Numerical} \Trule\Brule\\
%%%%%%%
%%%%%%%
Mechanical & $E_{\text{LLZO}}$ & 150\, GPa & \citeauthor{narayan2020modeling} (\cite{narayan2020modeling})
\Trule\Brule\\
& $\nu_{\text{LLZO}}$ & 0.26\, & \citeauthor{narayan2020modeling} (\cite{narayan2020modeling}) \Trule\Brule\\
& $E_{\text{Li}}$ & 4.91\, GPa & \citeauthor{narayan2018large} (\cite{narayan2018large}) \Trule\Brule\\
& $\nu_{\text{Li}}$ & 0.36 \, & \citeauthor{narayan2018large} (\cite{narayan2018large}) \Trule\Brule\\
Damage & $\Gamma$ & 0.04\, MPa$\cdot$s\, & \citeauthor{narayan2019gradient} (\cite{narayan2019gradient}) \Trule\Brule\\
& $\psicr \cdot \ell$ & 6.67\, J/m$^2$ & \citeauthor{narayan2020modeling} (\cite{narayan2020modeling}) \Trule\Brule\\
& $\ell$ & 5.0\, $\mu$m & Numerical 
\Trule\Brule\\
%%%%%%
\hline
\end{tabular}
\end{table}
We emphasize that the physics based thermodynamically consistent formulation \textit{allows virtually all material parameters to be found from the literature, either experimentally or from ab-initio simulations}. 
That is, the theoretical framework may be calibrated in a straightforward fashion and as such should prove useful in application to a number of engineering problems of relevance.

As detailed in Sect. \ref{free_enrg}, the parameter, $\psicr$ represents the dissipated energy per unit volume during the fracture process, while $\ell$ denotes a length scale to account for the damage gradient effects.
As discussed by Narayan and Anand \cite{narayan2019gradient} and De Borst and Verhoosel \cite{de2016gradient}, the gradient damage formulation will be mesh-independent, provided the element size, $h_e$, is sufficiently small compared to the length scale, $\ell$. Typically, an element size $h_e \le 0.2 \cdot \ell$ is considered sufficient.
Additionally, the product $\psicr \cdot \ell$ is related to the macroscopic critical energy release rate,  $G_c$ through $G_c \approx \psicr \cdot \ell$ \cite{anand2019modeling,mao2018fracture,talamini2018progressive}. 
Provided the length scale parameter, $\ell$ is chosen in a suitable physically realistic range, $\ell$ acts as an adjustable regularization parameter in our gradient damage theory.  
We prescribe a length scale, $\ell \approx 5 \, \mu$m to obtain a width for the fracture zone comparable in dimensions to published works in the literature \cite{yuan2021coupled,tantratian2021unraveling}. 
The characteristic element, size $h_e = 570\,\text{nm}$ is then chosen and satisfies the condition $h_e \le 0.2 \cdot \ell$ for mesh-independent results. This element size dictates the choice of a 350 x 350 finite-element mesh for the given $200\,\text{$\mu$m}$ x $200\,\text{$\mu$m}$ simulation domain. 
This ensures the interface is sufficiently narrow, while the number of elements required to resolve the interface does not become computationally intractable. 
With $\ell$ fixed, the value of $\psicr$ is then chosen such that $\psicr \cdot \ell = G_c = 6.67$ J/m$^2$ for LLZO.

\vspace{10pt}
\begin{remark}
To mitigate the generation of large stresses due to the elastic (rather than elastic-viscoplastic) model employed here for Li-metal, we  reduce the overall reaction-induced expansion associated with deposition of Li-metal within cracks.
We assume a 30\% expansion of Li-metal during electrodeposition in the $\mg$ direction.
That is, we choose $\Omega_\text{Li} \ximax = 0.3$ and with $\Omega_\text{Li} = 1.3\cdot10^{-5}$ m$^{3}$/mol known experimentally (cf. \citeauthor{tantratian2021unraveling} \cite{tantratian2021unraveling}), we use $\ximax = 2.31\cdot 10^4$ mol/m$^{3}$ in our simulations.
Theoretically, one should choose $\Omega_\text{Li} \ximax = 1$ (cf. Chen et al. \cite{chen2015modulation}).
We purposely use a smaller total expansion during electrodeposition to limit stress generation in Li-filaments. 
The value of $\Omega_\text{Li} \ximax = 0.3$ was calibrated to reproduce stresses within Li-filaments comparable to theoretical predictions based on linear elastic fracture mechanics models (cf. Klinsmann et al.\cite{klinsmann2019dendritic}).
\qed
\end{remark}

\begin{figure}[h!]
    \centering
    \includegraphics[width=6.5in]{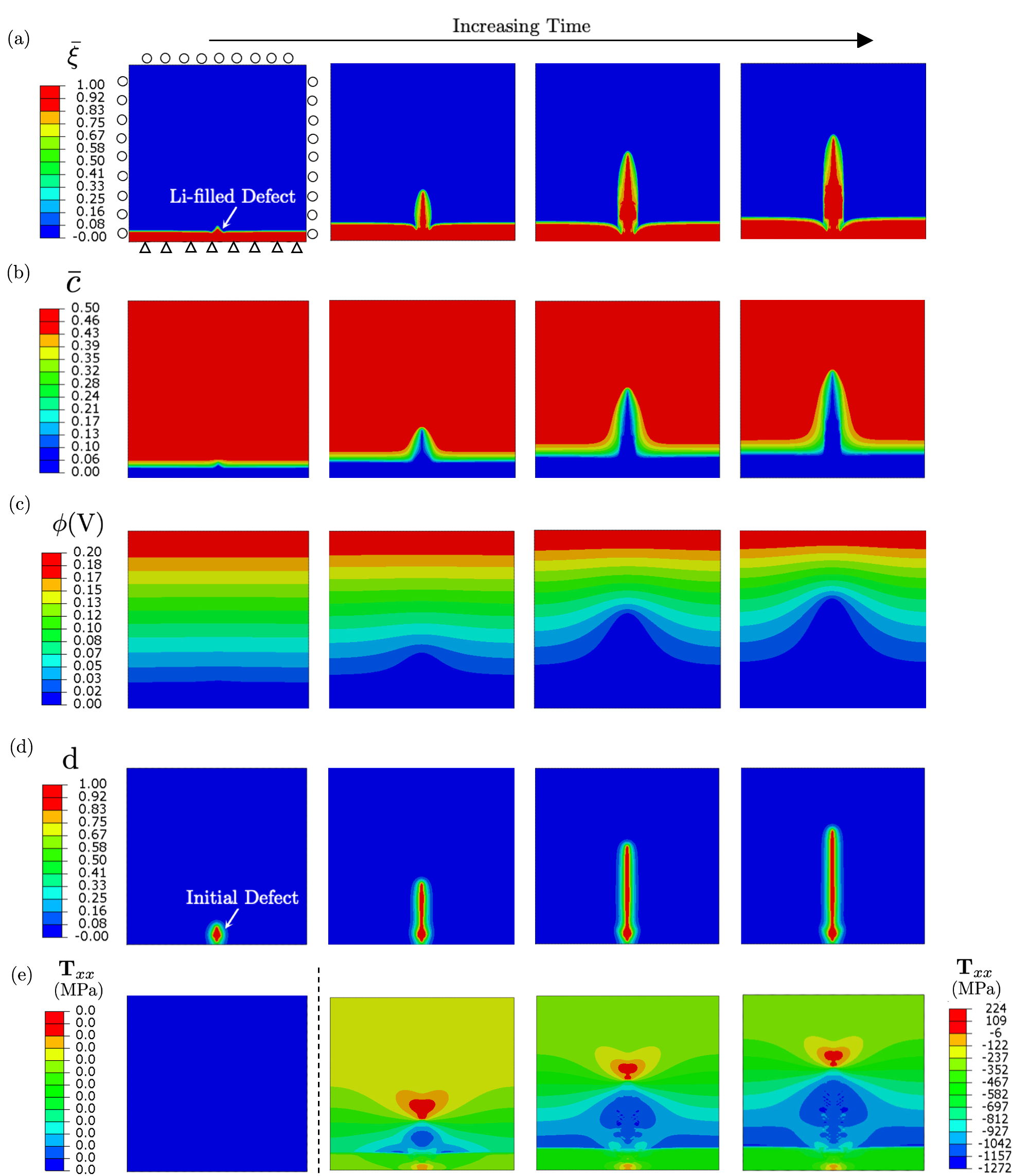}
    \caption{Evolution in time for (a) the extent of electrodepostion $\xibar$, (b) the normalized concentration $\cbar$, (c) the electric potential $\phi$(V), (d) damage $\dam$, and (e) the horizontal stress component $\bfT_{xx}$. The simulation is mechanically fully-constrained, with rollers present on the left and right edges.}
    \label{fig:rollers_profiles}
\end{figure}

The onset and evolution of the Li-metal filament with continuous fracture of the LLZO electrolyte is shown in Fig. \ref{fig:rollers_profiles}. 
For completeness, we illustrate the evolution of all five fields in time, namely from top to bottom (a) the extent of electrodeposition $\xibar$, (b) the normalized concentration $\cbar$, (c) the electric potential $\phi$, (d) damage $\dam$, and (e) the horizontal stress component, $\bfT_{xx}$. 
Note that the first column of Fig.~\ref{fig:rollers_profiles} shows the initial configuration of the Li-metal electrode-SSE assembly, where the domain is stress-free.
An animation of the simulation corresponding to Fig.
\ref{fig:rollers_profiles} is shown in the Supplemental Video {`S1\_FullyConstrained.mp4'}. 

Starting with a small imperfection at the anode-electrolyte interface, we can observe the growth in time of the amplitude of the Li-metal filament relative to the flat electrode sides, Fig. \ref{fig:rollers_profiles}(a). 
Fig. \ref{fig:rollers_profiles}(b-c) illustrate the evolution in normalized concentration and electric potential.
Critically, concentration of Li-ions remains unchanged in the bulk of the SSE, where the unaffected electrolyte (i.e. $ \cbar=0$ where $ \xibar=0 $) preserves a state of electroneutrality. 
In contrast, significant concentration gradients arise at the Li-Metal/SSE interface, where Li-ions are continuously consumed to plate fresh layers of Li-metal. 
Owing to the much higher conductivity of the Li-metal electrode compared to the SSE,  the electric potential across the freshly plated regions strictly follows the anode potential.
These profiles are consistent with works in the literature invoking a similar phase-field framework to model the growth of Li-metal filaments across SSEs \cite{tantratian2021unraveling,ren2015direct}.

Growth of a dominant Li-metal filament --- in contrast to uniform platting of Li-metal --- is attributed to localization of reaction kinetics at the tip of the protrusion, a phenomena governed by both electro-chemical and mechanical forces. 
From an electro-chemical standpoint, growth of Li-metal filaments is directly related to the concentration of Li-ions and electric potential at the reaction site, as described by the reaction driving force, $\mathcal{F}$ in \eqref{reac-evo}.
Fig.~\ref{fig:rollers_profiles}(b-c) illustrate the localization of larger concentration and electric potential gradients at the tip of the Li-metal filament, as it propagates from the metal anode towards the far electrolyte end. 
This results in an increase in driving force for electrodeposition, further localizing reaction kinetics and speeding up metal filament growth at these sites.

From a mechanical standpoint, Fig.~\ref{fig:rollers_profiles}(e) illustrates the evolution of the horizontal stress stress, $\textbf{T}_{xx}$ across the simulation domain.
Li-metal deposition inside the crack induces a large build-up in normal stresses acting on the crack surfaces. 
Owing to the stiffer nature of the solid electrolyte, Li-metal confined on the crack interior undergoes large compressive stresses, which in turn retard electrodeposition at the crack edges through the mechanically coupled reaction driving force $\calF$ in \eqref{reac-evo}.
The reverse is true at the tip of the crack, where large tensile stresses develop as the crack sides expand to accommodate the deposition of new lithium.
\textit{Tensile stresses at the crack tip facilitate electrodeposition of Li-metal, further driving filament formation, and importantly inducing successive damage of the solid state electrolyte}. 
As shown in Fig.~\ref{fig:rollers_profiles}(e), this state of stress is sustained as large tensile stresses drive crack propagation and in turn allow for further Li-metal platting inside the cracks, which sustains filament growth. 

Finally, Fig. \ref{fig:rollers_profiles}(d) illustrates the propagation of the electrodeposition-induced crack with simulation time. 
We emphasize that we do not predefine the crack path in our simulations. Instead, \textit{cracks are free to evolve in any arbitrary orientation, driven by a thermodynamically consistent driving force \eqref{dam_pde}}. 
Here, as there are no asymmetries in the simulation domain, the crack and subsequent Li-filament grow in a straight fashion from the anode through the electrolyte. The numerically simulated crack morphology is consistent with experimental observations of crack propagation in single-crystal SSE microstructures\cite{hao2021tracking,ning2021visualizing,swamy2018lithium}. 

\begin{figure}[h!]
    \centering
    \includegraphics[width=5in]{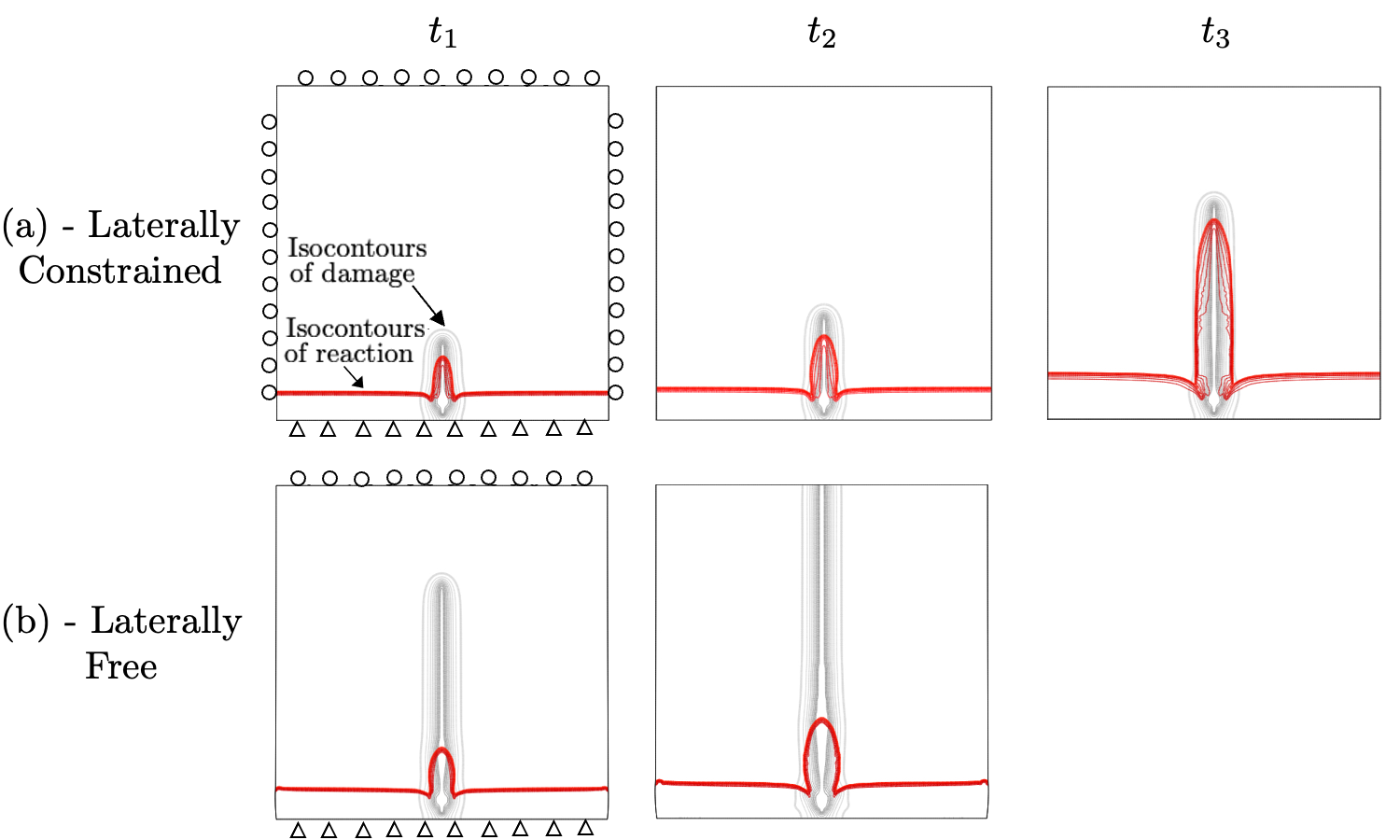}
    \caption{Evolution in time of the isocontours of damage $\dam$ (black lines) and reaction coordinate $\xibar$ (red lines) for (a - top row) a laterally constrained simulation, and (b - bottom row) a laterally free simulation.
    Comparing (a) and (b) demonstrates the critical role of mechanical confinement in dictating the degree to which the electrodeposition-induced crack becomes filled with Li-metal.
    }
    \label{fig:overlay_profiles}
\end{figure}

The importance of mechanical stress on the crack-electrodeposition interplay can be demonstrated by varying the degree of mechanical confinement to the solid-state architecture.
%late
Fig.~\ref{fig:overlay_profiles}  shows an
overlay of the isocontours of the damage field, d (black lines) and the reaction coordinate, $\xibar$ (red lines) for both a simulation with (a - top row) left and right sides of the domain constrained, i.e. "laterally constrained", and (b - bottom row) laterally free sides. 
The simulation results are shown at three synchronized times denoted $\{t_1,t_2,t_3\}$. We note here that the laterally free simulation terminates at $t_2$ when the crack reaches the top edge of the simulation domain, while the laterally constrained simulation continues. 
For completeness, contours of all five fields (as shown for the laterally constrained simulation in Fig.~\ref{fig:rollers_profiles}) are presented in the Supplemental Video {`S2\_LaterallyFree.mp4'} for the laterally free simulation.

We can clearly observe a significant difference in the extent of the Li-metal plated crack  when comparing the laterally constrained, Fig.~\ref{fig:overlay_profiles}(a), and laterally free, Fig.~\ref{fig:overlay_profiles}(b) simulations. 
In the laterally constrained case, the damage and extent of reaction fields evolve in a one-to-one fashion, with the crack remaining filled with Li-metal throughout. 
In contrast, \textit{the laterally free simulation predicts that the electrodeposition-induced cracks propagate significantly ahead of the growing Li-metal filament}.
In Fig.~\ref{fig:overlay_profiles}(b), we can observe the electrodeposition induced crack traversing the entire solid electrolyte domain, while the Li-metal filament trails behind. 
These results are in good qualitative agreement with the experimental observations of Ning et al. \cite{ning2021visualizing} and Hao et al. \cite{hao2021tracking}, where in-situ X-ray tomography revealed the propensity for Li-metal filaments to sometimes only partially fill cracked regions of the SSE, with cracks traveling at a much faster rate to reach the opposite cathode end plenty in advance of Li-deposits.
An animation of the simulation corresponding to Fig.
\ref{fig:overlay_profiles} is shown in the Supplemental Video {S3\_MechConfinement\_Overlay.mp4}.

Next, in Sect.  \ref{dam_mech_kinetics}, we demonstrate the manner in which microstructural features may be incorporated into our framework. Through this modeling, we elucidate the role of mechanics and damage on electrodeposition kinetics and resulting morphology of Li-metal filaments. Additionally, we demonstrate the utility of the framework by modeling the manner in which the morphology of Li-metal filaments changes for a LLZO microstructure depending on the relative fracture strength of the LLZO grains versus the grain boundaries.

%%%
%%%
%%%
\subsection{Modeling growth of Li-metal filaments in the presence of microstructural heterogeneity.}
\label{dam_mech_kinetics}

We present here a set of numerical simulations aimed to illustrate the manner in which grain boundaries may be modeled within the proposed framework.
Through these simulations, we elucidate the role of mechanics and damage on electrodeposition kinetics and resulting morphology of Li-metal filaments as they grow across a LLZO electrolyte. 
In Sect. \ref{transg_growth}, we considered only a homogeneous LLZO electrolyte, naturally resulting in the growth of a straight Li-metal filament through the SSE.

\begin{wrapfigure}[16]{r}{2.5in}
\vspace{-0.1in}
\includegraphics[width=2.5in]{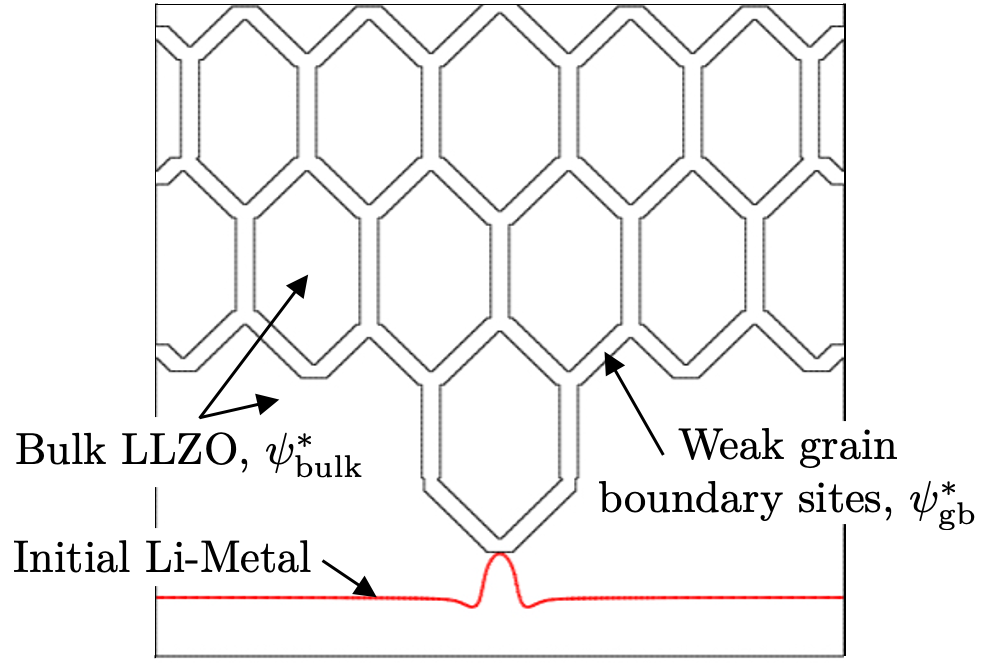}
\caption{Schematic of the simulation domain highlighting elements chosen to represent grain boundaries in the LLZO solid-state electrolyte.}
\label{fig:GBmesh}
\end{wrapfigure} 
We consider here a simple representation of a polycrystalline LLZO microstructure as schematically illustrated in Fig.~\ref{fig:GBmesh}.
The hexagonal pattern illustrated in Fig.~\ref{fig:GBmesh} shows the elements within the simulation domain, which represent the grain boundaries. 
A characteristic grain size of 30 $\mu$m is chosen for the design, consistent with the experimental reports of Sato et al. \cite{sato2020mapping} and Sharafi et al. \cite{sharafi2017controlling}. 
These grain boundary elements represent weaker structural sites characterized by a lower fracture energy, $\psicrgb$, compared to the bulk SSE, which is characterized by a fracture energy, $\psicrb$.
As such, grain boundaries are susceptible to damage and subsequent Li-metal electrodeposition \cite{ren2015direct}. All other mechanical properties are maintained the same for the entire simulation domain.

From an electro-chemical standpoint, we assume the grain boundaries to have the same electro-chemical properties (i.e. diffusivity, conductivity) as the grain bulk.
This assumption is purposely chosen to enable us to investigate {\it purely the mechanical} effects of grain boundaries on electrodeposition kinetics and resulting Li-metal filaments morphology.
As such, any effects sourcing from variation in electro-chemical properties between grain boundaries and the solid electrolyte bulk are neglected, although we note that these effects could also be incorporated in the theoretical framework presented. 
All material properties remain the same as those in Sect.~\ref{transg_growth} and listed in Tab.~\ref{tab:props}. 

Boundary conditions are identical to the fully-constrained simulation described in Sect.~\ref{transg_growth} and consistently here, an electrical potential, $\phi = 0.2\,\text{V}$ is applied to drive the system out of equilibrium.
We employ a similar configuration for the Li anode-SSE interface and introduce a small Li-filled defect. 
With Fig.~\ref{fig:GBmesh} in mind, we can observe that as damage and Li-metal filament grow across the SSE, they will eventually intersect with the mechanically weaker grain boundaries. 
At this point, Li-metal filaments can either propagate: i)  intergranularly across the mechanically weaker grain boundaries, ii) transgranularly by fracturing the bulk LLZO and continuing to grow in a straight line, or iii) through a combination of both modes. 
The growth direction and resulting morphology of Li-filaments is dictated by the thermodynamic reaction driving force \eqref{reac-evo}, which models the role of chemical, electrical and mechanical driving forces on Li-metal electrodeposition kinetics.

To illustrate the manner in which mechanics couples to electrodeposition kinetics, we perform a series of numerical simulations for the specific case of $\psicrgb = (1/3) \cdot \psicrb$.
To distinguish the two ways mechanics couples to electrodeposition, we consider first the reaction kinetics  \eqref{reac-evo}, which may be written as $\dot{\xibar} =  f_2(\dam) \hat{F}_1(\xibar,\calF)$. 
This highlights the first mechanism being the logistic function, $f_2(\dam)$ which directly couples the reaction kinetics to mechanical damage. 
As detailed in Sect.~\ref{reac_spec}, the function, $f_2(\dam)$ is introduced phenomenologically to restrict electrodeposition strictly within damaged regions across the SSE.  
Second, the electrodeposition driving force $\calF$ from \eqref{reac-evo} may be written as $\calF = \hat{F}_2(\cbar,\xibar,\phi) - \Ji h(\xibar) \Me \tendot \Ng$, where we now separate the stress-dependent term, which couples the reaction driving force to mechanical stresses.

\begin{figure}[h]
    \centering
    \includegraphics[width=6in]{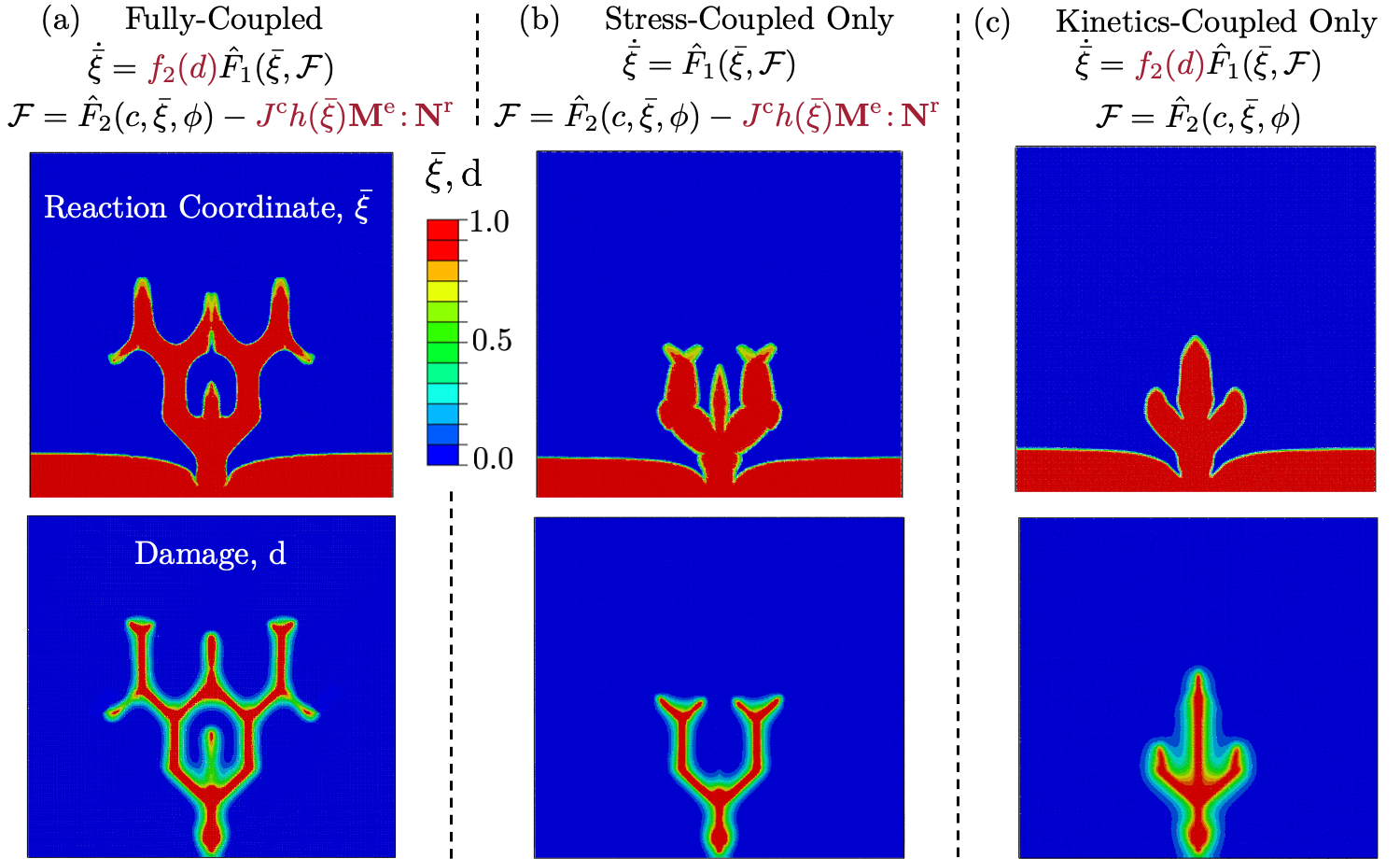}
    \caption{Contours of reaction coordinate $\xibar$ (top row) and damage $\dam$ (bottom row) for (a) a fully-coupled simulation, (b) a stress-coupled only simulation, and (c) a kinetics-coupled only simulation.}
    \label{fig:diff_couplings}
\end{figure}

Fig. \ref{fig:diff_couplings}(a) illustrates the fully-coupled simulation, where both mechanisms are active. 
We can observe the role of mechanically weaker grain boundaries in guiding the crack and subsequent growth of Li-metal filaments across the SSE. 
As shown in Fig.~\ref{fig:diff_couplings}(a), Li-metal filaments growth remains almost entirely confined within the damaged grain-boundaries, with a small vertical crack within the first grain appearing, but failing to continue to grow. 

Fig. \ref{fig:diff_couplings}(b) shows a numerical simulation in which the logistic function, $f_2(\dam)$ is removed, and direct coupling of electrodeposition kinetics to damage {\it is suppressed}. The only coupling between mechanics and electrodeposition is through the stress-coupled reaction driving force $\calF$.
Importantly, we can observe that even in the absence of a restriction preventing plating outside of cracks, {\it Li-metal filaments propagate primarily in cracked regions due solely to the stress-state within the cracks creating a favorable reaction driving force}.
Consistent with the results in Fig. \ref{fig:diff_couplings}(a), we observe the tendency for Li-metal to preferentially deposit across the mechanically weaker grain boundaries, even though a concurrent transgranular branch forms in this case.
This result illustrates that the presence of the logistic function, $f_2(\dam)$ in \eqref{xi_evolution_mod} does not significantly alter the resultant morphology of Li-metal filaments across the SSE microstructure.

Finally, Fig.~\ref{fig:diff_couplings}(c) illustrates the reverse coupling, where the stress-coupling of the reaction driving force $\calF$ {\it is suppressed}, while the reaction kinetics remain coupled to damage through the function $f_2(\dam)$.
In Fig.~\ref{fig:diff_couplings}(c), we observe an entirely different pattern of crack propagation and Li-metal deposition across the SSE. 
In contrast to simulation results with active stress-coupling, Figs.~\ref{fig:diff_couplings}(a,b), in which cracks and Li-metal filaments predominantly evolve across the weaker grain boundaries, \textit{the simulation with no stress-coupling predicts cracks and Li-metal filaments to grow primarily in a transgranular fashion}. 
That is, {\it in the absence of a stress-coupled reaction driving force $\calF$, the presence of weak grain boundaries is virtually unseen by the electrodeposition kinetics.}
This difference in pattern is expected, granted in the absence of stress-coupling, electrodeposition kinetics is dictated by purely electrochemical forces. 
As discussed in Sect.~\ref{transg_growth}, localization of higher concentration and electric potential gradients at the tip of the straight Li-metal branch compared to the slanted grain boundaries favors reaction and speeds up the growth of a dominant vertical Li-metal filament across the SSE.

These results highlight the importance of coupling the electrodeposition driving force, $\calF$ to mechanical stress.
Importantly, mechanical stresses in the presence of microstructural heterogeneities (i.e. grain boundaries, pores etc.) can significantly alter the morphological evolution of cracks and Li-metal filaments across the SSE. 
Such coupling has also been shown experimentally \cite{yu2017grain,cheng2017intergranular,ren2015direct}, and is reproduced here through the proposed theoretical framework. 

%%%%%%%%%%

Having discussed the manner in which mechanics couples to electrodeposition kinetics, we now turn our attention to the role of microstructure composition, specifically the role of grain boundaries with varying fracture energy. 
We consider a range of grain boundary fracture energies, $\psicrgb$, relative to the bulk solid electrolyte fracture energy, $\psicrb$, to explore the manner in which microstructural features with varying mechanical properties alter the morphology of cracks and Li-metal filaments across the SSE.

\begin{figure}[h]
    \centering
    \includegraphics[width=6in]{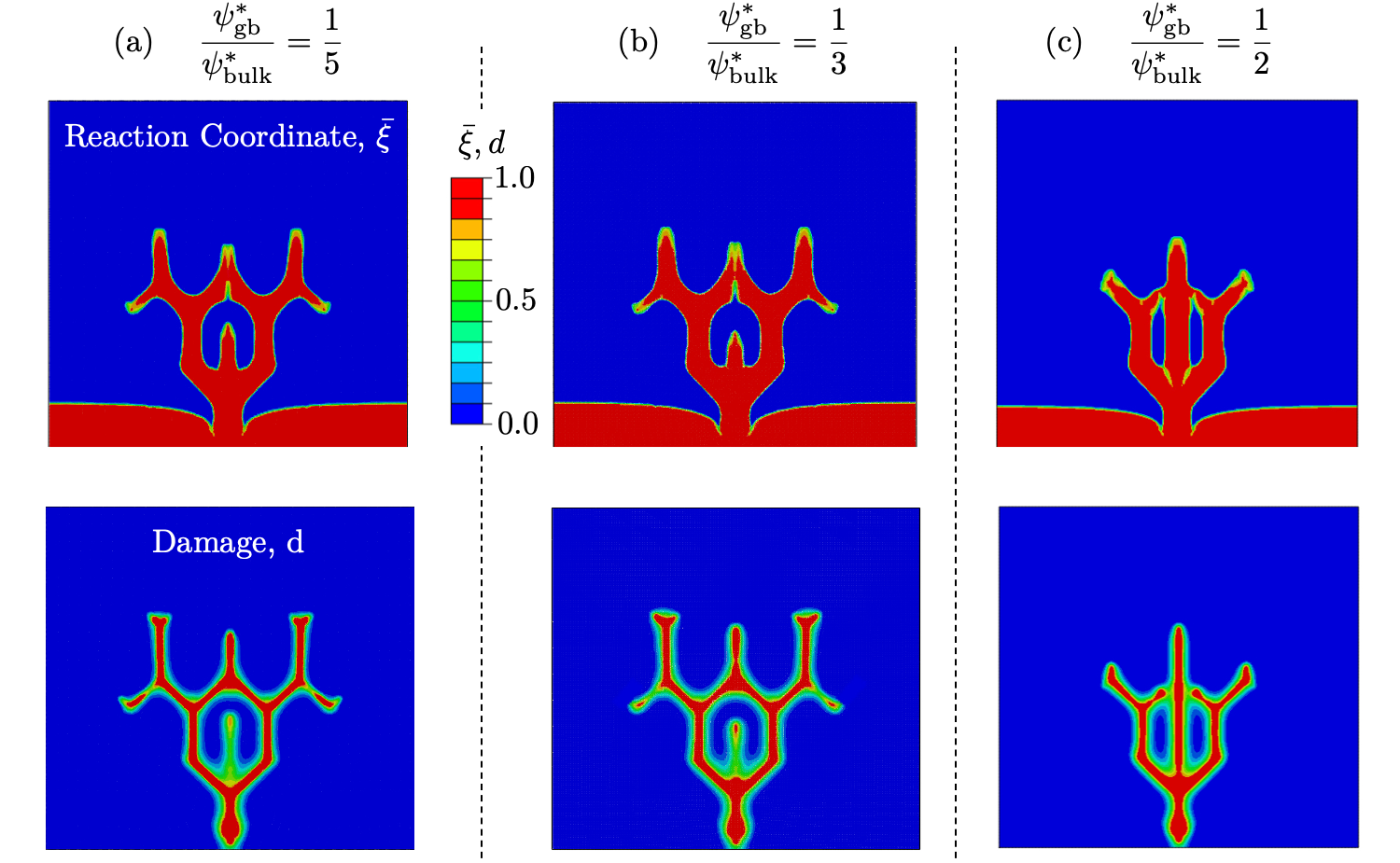}
    \caption{
    Contours of reaction coordinate $\xibar$ (top row) and damage $\dam$ (bottom row) with varying grain boundary fracture energies, respectively (a)  $\psicrgb = (1/5) \cdot \psicrb$, (b) $\psicrgb = (1/3) \cdot \psicrb$, and (c) $\psicrgb = (1/2) \cdot \psicrb$. }
    \label{fig:gb_var_fract}
\end{figure}

Fig.~\ref{fig:gb_var_fract} shows contours of reaction coordinate (top row) and damage (bottom row) for simulations with (a) $\psicrgb = (1/5) \cdot \psicrb$, (b) $\psicrgb = (1/3) \cdot \psicrb$, and $\psicrgb = (1/2) \cdot \psicrb$.
Note that all these simulations are fully-coupled, and thus the result in Fig.~\ref{fig:gb_var_fract}(b) is identical to that in Fig.~\ref{fig:diff_couplings}(a).
An animation of the simulations corresponding to Fig. \ref{fig:gb_var_fract} is shown in the Supplemental Video {S4\_GB\_Filament\_Growth.mp4}.

For both cases with $\psicrgb = (1/5) \cdot \psicrb$ and $\psicrgb = (1/3) \cdot \psicrb$, Figs.~\ref{fig:gb_var_fract}(a,b), we can observe the role of weaker grain boundaries in guiding Li-metal to preferentially deposit at these sites. 
Owing to their lower fracture energy, grain boundaries are easier to damage compared to the bulk SSE. This creates a favorable stress-state within the fractured grain boundaries, making it favorable for growth of Li-metal filaments to follow the grain boundary network, suppressing the growth of a vertical transgranular crack and Li-metal branch.
These results are consistent with experimental observations, which report Li-metal filaments to preferentially grow along the grain boundaries (i.e. integranularly) in a polycrystalline LLZO electrolyte \cite{ren2015direct,cheng2017intergranular}.

With increase in fracture energy of the grain boundaries relative to the bulk SSE, $\psicrgb = (1/2) \cdot \psicrb$ in Fig.~\ref{fig:gb_var_fract}(c), we observe a change in the resulting crack and Li-metal filament morphology.
We observe now the formation of a distinct vertical transgranular branch, propagating through the SSE in advance of concurrent Li-metal growth proceeding across the grain boundaries.  
As the fracture energy of the grain boundaries increases closer to that of the electrolyte bulk, fracturing these sites becomes more difficult, and transgranular growth becomes preferable.
This result suggests there exists a threshold fracture energy ratio, where the electrochemical driving force overcomes the weak grain boundaries and the morphology of Li-metal filaments growth  transitions from primarily intergranular to primarily transgranular. 
Clearly, the limiting case of $\psicrgb = \psicrb$ is identical to the simulation results shown in Fig.~\ref{fig:rollers_profiles} and produces a single vertical Li-filament.

At present, a microstructure description of the process of Li-metal filament growth across SSEs is missing and necessary.
The theoretical framework summarized and specialized in Sects.~\ref{sect:summ_general} and \ref{theory_spec} and numerical simulations presented here show the first framework capable of predicting the resulting morphology of Li-metal filaments for a given SSE architecture, including a transition from primarily intergranular to primarily transgranular filament growth.
%

%%%
%%%
%%%
\section{Concluding Remarks}
\label{conclusion}

We have formulated  and numerically implemented a thermodynamically-consistent electro-chemo-mechanical gradient theory, which couples electrochemical reactions with mechanical deformation and damage in solids. 
The framework models transport of ionic species across the solid host under diffusion/migration mechanisms and concurrent electrochemical reaction at cracks across the solid host, where ionic species are reduced to deposit a new compound. 
Critically, the theory captures the interplay between growth-induced fracture of the solid host and electrodeposition of a new material inside cracks by tracking the damage and extent of electrodeposition using separate phase-field variables.

While the framework is general in nature, a specialization towards modeling of Li-metal filament growth across SSEs for application in all-solid-state batteries is demonstrated. 
In particular, we elucidate the coupling of electrical, chemical, and mechanical processes, which govern the growth of Li-metal filaments across SSEs with and without account for microstructure heterogeneity.
Consistent with experiments, we demonstrate the ability of the framework to numerically reproduce both transgranular and intergranular growth of metal protrusions across the SSE microstructure.

Major contributions of this work include:
\begin{itemize}
    %
    %%%
    \item 
    The framework enables for modeling concurrent diffusion, reaction, deformation and damage in solids. Importantly, the theory extends the current literature in continuum mechanics to model solids undergoing electrochemical reactions, which are intricately coupled to damage of the underlying solid host. 

    The interplay between electrodeposition of new material inside damaged (fractured) regions is modeled. Consistent with experiments, electrodeposition is restricted within damaged regions across the solid conductor, which supply the necessary vacant space for accommodating a new material. 
    This feature provides an extension to current electro-chemo-mechanical phase-field formulations in the literature, which do not model growth-induced fracture of SSE due to Li-deposition and indistinguishably allow for Li-metal plating across the host.
   %
    %
    %%%
    \item  
    The thermodynamically consistent treatment enables derivation of a physically-motivated reaction driving force in which different contributions of energy, configuration entropy, electrostatics, mechanical deformations and damage can be readily identified. 
    Particularly useful then is the fact that material properties driving the electrodeposition kinetics can be calibrated from the literature or experiments. 
    %
    %%%
    \item 
    The framework elucidates the role of mechanical confinement on the crack-electrodeposition morphology. 
    Under specific mechanical boundary conditions, we numerically reproduce the presence of cracks, which are only partially filled with Li-metal. 
    %
    %
    %%%
    \item 
    Finally, we demonstrate the role of microstructural heterogeneities, in particular grain boundaries, in dictating the morphology of cracks and Li-metal filaments across the SSE microstructure.
    Both transgranular and intergranular growth mechanisms are numerically reproduced consistent with experiments, elucidating the critical role of mechanics on electrodeposition kinetics and the resulting morphology of Li-metal filaments across the SSE.
\end{itemize}

Growth of Li-metal filaments across SSEs constitutes a critical degradation mechanism, hindering commercialization of next-generation all-solid-state batteries. 
It is thus critical, from both an experimental and modeling perspective, to understand the interplay between non-uniform electrodeposition kinetics and the SSE microstructure composition.
In particular, the manner in which size and distribution of defects and microstructural heterogeneities ultimately govern the growth of Li-metal filaments. 
The proposed framework, in conjunction with experiments, enhances our current understanding of the electro-chemo-mechanical processes 
, which govern the initiation and growth of Li-metal filaments across SSEs. 
From a design standpoint, this work provides then a utilitarian numerical tool, which can help elucidate new strategies to mitigate growth of Li-metal filaments and enable for successful commercialization of all-solid-state batteries.

\section*{CRediT authorship contribution statement}
Donald Bistri: Design and implementation of the research, Analysis of results, Writing of the manuscript. 
Claudio V. Di Leo:
Design and implementation of the research, Analysis of results, Writing of the manuscript.

\section*{Acknowledgements}

Support is acknowledged from NASA Grant Number 80NSSC21M0101.

\clearpage

%%%%%%%%%%%%%%%%%%%%%%%%%%%%%%%%%%%%%%%%%%%
%%%%%%%%%%%%%%%%%%%%%%%%%%%%%%%%%%%%%%%%%%%
%%%%%%%%%%%%%%%%%%%%%%%%%%%%%%%%%%%%%%%%%%%
\appendix

\section*{Appendix A - Principle of Virtual Work, Macro- and Microforce Balance Laws}
\label{appendixA}
We present here a detailed derivation of the macro- and microforce balance laws for the rate-like kinematical descriptors in our theory. We consider a list of generalized virtual velocity fields to be given by $\calV = ( \delta \y, \delta \Fe,  \delta \xir, \nabla \delta \xir,\delta \dam, \nabla \delta \dam )$ and constrained through the kinematic relation \eqref{Defconstraint_vir}  presented in Sect. \ref{sect:macro_micro_vir_pow}. 
%%%%%%%%%
For any part $\mathcal{P}$ with outward unit normal $\bfn_\mat$ of the reference body $\mathcal{B}$, the internal and external power are given by
\setcounter{equation}{0}
\renewcommand{\theequation}{A.\arabic{equation}}
%%%%%%%%%
\begin{equation}
\begin{split}
\delta W_\text{ext} (P,\calV) &= \int_{\partial\p} \bft_\mat(\bfn_\mat) \cdot \delta  \y \, da_\mat 
%%%%
+  \int_{\p} \bfb_\mat \cdot \delta  \y  \, dv_\mat 	
%%%%%
+ \int_{\partial \p} \eta \delta \xir \, da_\mat 
%%%%%
+ \int_{\partial \p} \gamma \delta \dam \, da_\mat \\[4pt]
%%%%%%%%%
%%%%%%%%%
\delta W_\text{int} (P,\calV) &= \int_{\p} (
\Se \tendot \delta \Fe
%%%%%%%%
+ E  \delta \xir
%%%%%%
+\bfG \cdot \nabla \delta \xir 
%%%%%%%
+ \varpi  \delta \dam 
%%%%%%%%
+ \bfzeta \cdot \nabla \delta \dam ) \, dv_\mat 
\end{split}
\label{eq:VirtPower_app}
\end{equation}
%%%%%%%%
To derive the balance laws, we invoke the power balance requirement (c.f. Sect. \ref{sect:macro_micro_vir_pow}) and demand that $ \delta W_\text{ext} (P,\calV) =  \delta W_\text{int} (P,\calV)$ for all generalized virtual velocities $\calV$. 
%%%%%%%

First, we derive the local macroforce balance. Let $ \delta \xir =0, \nabla \delta \xir = \mathbf{0}, \delta\dam = 0, \nabla\delta\dam = \mathbf{0} $ such that the kinematic constraint \eqref{Defconstraint_vir} yields the relation  $\delta \Fe = (\nabla \delta \y) \bfF^{-1} \Fe$.
%%%%%%%
For this choice, the principle of virtual power gives,
%%%%%%%%%
\begin{equation}
\int_{\partial\p} \bft_\mat(\bfn_\mat) \cdot \delta  \y  \, da_\mat +  \int_{\p} \bfb_\mat \cdot \delta  \y \,dv_\mat=\int_{\p} \Se \tendot \delta \Fe \, dv_\mat =  \int_{\p} \Se \Fcit \tendot  \nabla \delta \y  \,dv_\mat
\label{balance1}
\end{equation}
%%%%%%%%%
which by defining 
%%%%%%%%%
\begin{equation}
\T_\mat \Def  \Se  \Fcit 
\label{eq:stress1}
\end{equation}
%%%%%%%%
and applying the divergence theorem on \eqref{balance1} leads to the  macroscopic force balance
%%%%%%%%
\begin{equation} 
\text{Div} \, \T_\mat + \bfb_\mat = 0,
\quad 
\text{and the traction boundary condition}
\quad
\bft_\mat(\bfn_\mat) = \T_\mat \bfn_\mat.
\label{balance2}
\end{equation}
%%%%%%%%%%%%
As is standard, the Piola stress $\T_\mat$ is related to the symmetric Cauchy stress $\T$ through
%%%%%%%%%%%
\begin{equation}
\T_\mat = J \T \bfF^{-\trans},
\label{eq:cauchy4}
\end{equation}
%%%%%%%%%
and for future use we may write  $\Se = J \T \Feit$.
%%%%%%%

Subsequently, we derive the local microforce balances. The first microforce balance associated with reaction is obtained by letting $ \delta \y=\mathbf{0}, \delta \dam =0, \nabla\delta \dam =\mathbf{0}$, such that \eqref{Defconstraint_vir} yields,
%%%%%%%%
\begin{equation}
\begin{split}
\Se \tendot \delta \Fe &= \Se \tendot \Big (- h(\xibar) \Fe \Ng  \Big ) \delta \xir \\
&=\Big ( -  \Fet \Se \tendot \big(h(\xibar) \Ng \big) \Big ) \delta \xir \\
& = \Big (- (J \Fet \T \Feit  ) \tendot \big(h(\xibar) \Ng \big) \Big ) \delta \xir \\
&  =\Big(  -\Jc h(\xibar) \Me \tendot \Ng \Big) \delta \xir
\label{reac_stess_power}
\end{split}
\end{equation}
%%%%%%%%
where we have defined the elastic Mandel stress as,
%%%%%%%%%
\begin{equation}
\Mme \Def \Je \Fet \T \Feit  
\label{eq:Mandel_app}
\end{equation}
%%%%%%%
For this choice, and accounting for (A.6), the virtual power balance then yields
%%%%%%%%%
\begin{equation}
\int_{\partial \p} \eta \delta \xir  da_\mat 
= \int_{\p} \left( -\Jc  h(\xibar) \Me \tendot \Ng \delta \xir +  E \delta \xir + \bfG \cdot \nabla \delta \xir \right) dv_\mat.
\end{equation}
%%%%%%%%%
Applying the divergence theorem leads to
%%%%%%%%%
\begin{equation}
\int_{\p} \left(  -\Jc  h(\xibar) \Me \tendot \Ng \delta \xir +  E \delta \xir - \text{Div} \bfG \, \delta \xir \right)dv_\mat =  \int_{\partial \p} (\eta-\bfG \cdot \bfn_\mat) \delta \xir  da_\mat
\end{equation}
%%%%%%%%%
Granted this relationship must hold for all $\mathcal{P}$ and all values of $\delta \xir$, standard variational arguments yield the second microforce balance,
%%%%%%%%%
\begin{equation}
E -  \Jc h(\xibar) \Me \tendot \Ng  -  \text{Div} \bfG = 0
\label{micro2_app}
\end{equation}
%%%%%%%%%%
along with the corresponding boundary condition
%%%%%%%%%%
\begin{equation}
\eta(\bfn_\mat) = \bfG \cdot \bfn_\mat.
\label{eq:mirco2_BC_app}
\end{equation}
%%%%%%%%%
Lastly, we derive the local microforce balance associated with damage. Let  $\delta \y=\mathbf{0},  \delta\xir= 0$ and $\nabla\delta\xir = \mathbf{0}$. In a similar fashion, the virtual power balance statement then yields
%%%%%%%%%
\begin{equation}
\int_{\partial \p} \gamma \delta \dam  \, da_\mat 
=  \int_{\p} \left( \varpi\delta \dam + \bfzeta\cdot \nabla\delta \dam \right) dv_\mat
\label{bal_dam}
\end{equation}
%%%%%%%%%
Applying the divergence theorem on \eqref{bal_dam} gives,
\begin{equation}
\int_{ \p} (\text{Div}\bfzeta-\varpi) \delta \dam \, dv_\mat + \int_{\partial\p} \left(\gamma - \bfzeta \cdot  \bfn_\mat \right)\delta \dam \, da_\mat  = 0
\end{equation}
%%%%%%%%
which must hold for all $\mathcal{P}$ and all values of $\delta \dam$, yielding the second microforce balance 
%%%%%%%%%
\begin{equation}
\text{Div}\bfzeta - \varpi = 0
\label{eq:micro2_app}
\end{equation}
%%%%%%%%
along with the corresponding boundary condition
%%%%%%%%
\begin{equation}
\gamma(\bfn_\mat) = \bfzeta \cdot  \bfn_\mat
\label{eq:mirco2_BC_app}
\end{equation}
%%%%%%%%%%
In summary, using the principle of virtual work we have derived one macroforce balance for the Piola stress, $\T_\mat$ as well as two corresponding microforce balances for the stresses
$\{E,\bfG, \bfzeta, \varpi \}$ as outlined in Sect. \ref{sect:macro_micro_vir_pow}.

\clearpage
%%%%%%%%%%%%%%%%%%%%%%%%%%%%%%%%%%%%%%%%%%%%%%%%%%%%%%%%%%%%%%%%%%%%%%%%%%%%%%%%%%%%%
%                           references
%%%%%%%%%%%%%%%%%%%%%%%%%%%%%%%%%%%%%%%%%%%%%%%%%%%%%%%%%%%%%%%%%%%%%%%%%%%%%%%%%%%%
\setlength\bibitemsep{\baselineskip}  

\printbibliography[title={References}]

\end{document}